\documentclass[aps,prl,twocolumn,nofootinbib,longbibliography,amsfonts,amssymb,amsmath,superscriptaddress,floatfix]{revtex4-2}

\usepackage[utf8]{inputenc}
\usepackage{graphicx}
\usepackage[caption=false]{subfig}
\usepackage{physics}
\usepackage{xspace}
\usepackage{mathtools} 
\usepackage[dvipsnames]{xcolor}
\usepackage[USenglish]{babel}
\usepackage{microtype}
\usepackage{appendix}
\usepackage{makecell}
\usepackage[normalem]{ulem}

\definecolor{okblue}{HTML}{3A6FF7}   % L=70, C=0.15, h=255
\definecolor{okcyan}{HTML}{20B8D6}   % L=70, C=0.15, h=220
\definecolor{okgreen}{HTML}{43C06B}  % L=70, C=0.15, h=150
\definecolor{okpurple}{HTML}{A05FE8} % L=70, C=0.15, h=315
\definecolor{okamber}{HTML}{D58A32}  % L=70, C=0.15, h=70

\definecolor{okblueSoft}{HTML}{B9CCFF} 
\definecolor{okcyanSoft}{HTML}{B7EAF3}
\definecolor{okgreenSoft}{HTML}{C6F0D3}
\definecolor{okpurpleSoft}{HTML}{E5D1FA}
\definecolor{okamberSoft}{HTML}{F2D7B8}

\usepackage{tikz}
\usetikzlibrary{matrix}
\usetikzlibrary{shapes,arrows.meta,positioning,graphs,fit,backgrounds,calc,shapes.geometric}
\usetikzlibrary{decorations.pathreplacing}
\tikzset{
  every path/.style={line width=0.7pt},
  >={Latex[scale=.6]},
  wide/.style={minimum width=55mm},
  node distance=4mm,
  every node/.style={font=\footnotesize},
  imgbox/.style={rectangle,inner sep=0},
  box/.style={draw, rounded corners, align=center, inner sep=3pt, minimum height=6mm, minimum width=3mm},
  dist/.style={box, fill=okblueSoft, draw=okblue},                           % sampling / distributions
  nn/.style={box, fill=okamberSoft, draw=okamber},      % learned modules
  data/.style={box, fill=okcyanSoft, draw=okcyan},                     % data collections
  realdata/.style={box, fill=okpurpleSoft, draw=okpurple},
  decision/.style={diamond, draw=okgreen, aspect=1.6, inner sep=1pt, align=center, font=\scriptsize, text width=10mm, fill=okgreenSoft},
  group/.style={draw=black!60, dashed, rounded corners, inner sep=4pt},
  note/.style={font=\scriptsize, inner sep=1pt},
  edge/.style={->, draw=black!70}
}
\tikzset{
  textbox/.style={
    align=center,
    draw=none,
    fill=none,
    inner sep=0pt,
    minimum width=0pt,
    minimum height=0pt,
  },
}
\newcommand{\hyperparams}{\Lambda}

\newcommand{\p}{p}
\newcommand{\hu}{\,{\rm km \,s^{-1} \, Mpc^{-1}}} %
\newcommand{\msun}{M_\odot}

\newcommand{\icarogw}{\texttt{icarogw}}

\newcommand{\Data}{\mathcal{D}}

\newcommand{\ppop}{p_{\rm pop}(\theta\mid\hyperparams)}

\newcommand{\ms}{m_{{\rm s}}}
\newcommand{\md}{m_{{\rm d}}}
\newcommand{\msone}{m_{1,{\rm s}}}
\newcommand{\mstwo}{m_{2,{\rm s}}}
\newcommand{\mdone}{m_{1,{\rm d}}}
\newcommand{\mdtwo}{m_{2,{\rm d}}}

\newcommand{\widetildeparams}{\widetilde{\theta}}

\newcommand{\embedding}{\mathcal{Z}}
\newcommand{\mmin}{m_\mathrm{min}}
\newcommand{\mmax}{m_\mathrm{max}}

\newcommand{\plp}[0]{\textsc{Power Law + Peak}\xspace}
\newcommand{\dingoP}[0]{\textsc{Dingo-Pop}\xspace}

\newcommand{\dingo}[0]{\textsc{Dingo}\xspace}

\usepackage{xfp}
\usepackage{siunitx}
\newcommand{\trainepochsimple}{800}
\newcommand{\batchsizesimple}{128}
\newcommand{\populationsperepochsimple}{50000} % no comma
\newcommand{\populationsizeminsimple}{25}
\newcommand{\populationsizemaxsimple}{1000}
\newcommand{\totalsamplesSci}{%
  \num{\fpeval{\trainepochsimple * \populationsperepochsimple *
               (\populationsizemaxsimple + \populationsizeminsimple)/2}}%
}

\newcommand{\totalpopulationsSci}{%
  \num{\fpeval{\trainepochsimple * \populationsperepochsimple}}
}

\newcommand{\estimatepdet}{\hat p(\mdone, \mdtwo, d_L)}

\newcommand{\mug}{\mu_{\rm g}}
\newcommand{\sigmag}{\sigma_{\rm g}}
\usepackage[acronym, toc, nonumberlist]{glossaries}

\glsdisablehyper
\setacronymstyle{long-short}

\newacronym{ns}{NS}{neutron star}
\newacronym{bh}{BH}{black hole}
\newacronym{bbh}{BBH}{binary black hole}
\newacronym{bns}{BNS}{binary neutron star}
\newacronym{nsbh}{NSBH}{neutron star black hole}

\newacronym{eos}{EoS}{equation of state}
\newacronym{gw}{GW}{gravitational wave}
\newacronym{gr}{GR}{general relativity}
\newacronym{snr}{SNR}{signal-to-noise ratio}

\newacronym{lisa}{LISA}{Laser Interferometer Space Antenna }
\newacronym{ligo}{LIGO}{Laser Interferometer Gravitational wave Observatory}
\newacronym{kagra}{KAGRA}{KAmioka GRavitational wave detector}
\newacronym{eob}{EOB}{effective one-body}
\newacronym{em}{EM}{electromagnetic}
\newacronym{lcdm}{$\Lambda$CDM}{$\Lambda$ cold dark matter}
\newacronym{pl}{PL}{power law}
\newacronym{plg}{PLG}{power law and Gaussian}
\newacronym{kde}{KDE}{kernel density estimate}
\newacronym{de}{DE}{dark energy}
\newacronym{cdf}{CDF}{cumulative density function}

\newacronym{lvk}{LVK}{LIGO-Virgo-KAGRA}
\newacronym{ego}{EGO}{European gravitational observatory}
\newacronym{asd}{ASD}{amplitude spectral density}
\newacronym{psd}{PSD}{power spectral density}
\newacronym{mcmc}{MCMC}{Monte Carlo Markov chain}
\newacronym{hlv}{HLV}{Hanford Livingston Virgo}
\newacronym{pe}{PE}{parameter estimation}
\newacronym{cbc}{CBC}{compact binary coalescence}
\newacronym{aligo}{aLIGO}{advanced LIGO}
\newacronym{far}{FAR}{false alarm rate}

\newacronym{cl}{CL}{confidence level}

\newacronym{pn}{PN}{post-Newtonian}
\newacronym{nr}{NR}{numerical relativity}
\newacronym{ppisn}{PPISN}{pulsation pair-instability supernova}
\newacronym{pisn}{PISN}{pair instability-supernova}
\newacronym{et}{ET}{Einstein Telescope}
\newacronym{ce}{CE}{Cosmic Explorer}

\newacronym{cmb}{CMB}{cosmic microwave background}
\newacronym{lss}{LSS}{large scale structure}
\newacronym{isco}{ISCO}{innermost stable orbit}

\newacronym{oi}{Oi}{observation run $i$}
\newacronym{gwtci}{GWTC-i}{gravitational wave transient catalog $i$}

\newacronym{2g}{2G}{second generation}
\newacronym{3g}{3G}{third generation}

\newacronym{bao}{BAO}{baryonic acoustic oscillation}

\newacronym{wkb}{WKB}{Wentzel–Kramers–Brillouin}

\newacronym{dpg}{DPG}{Dvali-Gabadadze-Porrati}
\newacronym{dhost}{DHOST}{degenerate higher-order scalar-tensor}
\newacronym{mg}{MG}{modified gravity}
\newacronym{des}{DES}{dark energy survey}

\newacronym{tt}{TT}{transverse-traceless}

\newacronym{sgwb}{SGWB}{stochastic gravitational wave background}
\newacronym{dgrb}{DGRB}{diffuse $\gamma$-ray background}
\newacronym{gut}{GUT}{grand unified theory}

\newacronym{ng}{NG}{Nambu-Goto}
\newacronym{gbr}{GBR}{gravitational backreaction}

\newacronym{nf}{NF}{normalizing flow}
\newacronym{ml}{ML}{machine learning}
\newacronym{lfi}{LFI}{likelihood-free inference}
\newacronym{nn}{NN}{neural network}
\newacronym{dingo}{DINGO}{deep inference for gravitational wave observations}
\newacronym{gpu}{GPU}{graphics processing unit}
\newacronym{hba}{HBA}{hierarchical Bayesian analysis}

\newacronym{kl}{KL}{Kullback-Leibler}
\newacronym{js}{JS}{Jensen-Shannon}
\newacronym{ks}{KS}{Kolmogorov–Smirnov}

\newacronym{smbh}{SMBH}{supermassive black hole}
\newacronym{agn}{AGN}{active galactic nuclei}
\newacronym{spa}{SPA}{stationary phase approximation}

\newacronym{pta}{PTA}{pulsar timing array}

\newacronym{npe}{NPE}{neural posterior estimation}

\newacronym{dm}{DM}{dark matter}

\newacronym{grf}{GRF}{Gaussian random field}

\newacronym{desi}{DESI}{Dark Energy Spectroscopic Instrument}

\newacronym{sbi}{SBI}{simulation-based inference}

\renewcommand{\sec}[1]{\emph{#1.---}}

\usepackage[pdftex]{hyperref}

\begin{document}

\title{End-to-End Population Inference from Gravitational-Wave Strain using Transformers}

\author{Konstantin Leyde}
\email{kleyde@flatironinstitute.org}
\affiliation{Center for Computational Astrophysics, Flatiron Institute, 162 5th Ave, New York, NY 10010}
\affiliation{Institute of Cosmology and Gravitation, University of Portsmouth, \\
Burnaby Road, Portsmouth PO1 3FX, United Kingdom}

\author{Stephen R. Green}
\email{stephen.green2@nottingham.ac.uk}
\affiliation{Nottingham Centre of Gravity \& School of Mathematical Sciences, University of Nottingham, University Park, Nottingham NG7 2RD, United Kingdom}

\author{Maximilian Dax}
\email{maximilian.dax@tuebingen.mpg.de}
\affiliation{ELLIS Institute T\"ubingen, Max Planck Institute for Intelligent Systems, T\"ubingen AI Center, T\"ubingen, Germany}

\author{Matthew Mould}
\email{matthew.mould@nottingham.ac.uk}
\affiliation{Nottingham Centre of Gravity \& School of Mathematical Sciences, University of Nottingham, University Park, Nottingham NG7 2RD, United Kingdom}

\author{Cecilia Maria Fabbri}
\email{cecilia.fabbri@nottingham.ac.uk}
\affiliation{Nottingham Centre of Gravity \& School of Mathematical Sciences, University of Nottingham, University Park, Nottingham NG7 2RD, United Kingdom}

\author{Jonathan Gair}
\email{jonathan.gair@aei.mpg.de}
\affiliation{Max Planck Institute for Gravitational Physics (Albert Einstein Institute)\\
  Am M\"uhlenberg 1, 14476 Potsdam, Germany}

%Collaboration name if desired (requires use of superscriptaddress
%option in \documentclass). \noaffiliation is required (may also be
%used with the \author command).
%\collaboration can be followed by \email, \homepage, \thanks as well.
%\collaboration{}
%\noaffiliation

\date{\today}

\newcommand{\kcom}[1]{{\color{magenta} K: #1}}
\newcommand{\scom}[1]{{\color{blue} S: #1}}
\newcommand{\mcom}[1]{{\color{red} M: #1}}
\newcommand{\ccom}[1]{{\color{green} C: #1}}
\renewcommand{\mm}[1]{{\color{cyan}MM: #1}}

\begin{abstract}
The population of compact binaries encodes information about their astrophysical origins and the expansion of the universe. Hierarchical Bayesian methods infer these properties by combining single-event posteriors. As catalogs grow, however, this approach becomes computationally expensive and is subject to increasing Monte Carlo uncertainty. We introduce \dingoP, a simulation-based framework that infers population posteriors directly from gravitational-wave strain data. The data for each event are embedded into low-dimensional tokens and combined using a transformer trained on simulated catalogs subject to selection effects.
This enables (i) population inference without per-event Monte Carlo sampling noise, (ii) amortization across variable catalog sizes using a single network, and (iii) end-to-end inference in about one second. 
We train a network for catalog sizes of 25 to 1000 events, and obtain  well-calibrated posteriors consistent with traditional methods. By avoiding per-event analyses that can take hours to days, \dingoP enables new classes of large-scale injection studies; as an application, we examine how spectral-siren Hubble constant uncertainties change with catalog size. 
\end{abstract}

\maketitle

\sec{Introduction}Population inference is now central to gravitational-wave (GW) astronomy. The latest catalog, GWTC-4.0~\cite{LIGOScientific:2025hdt,LIGOScientific:2025slb}, contains 218 events observed by the LIGO–Virgo–KAGRA Collaboration (LVK)~\cite{LIGOScientific:2014pky,det4-VIRGO:2014yos,det6-KAGRA:2020tym}. By combining information across events, one can measure merger rates and source distributions~\cite{LIGOScientific:2025pvj}, perform tests of general relativity~\cite{LIGOScientific:2021sio}, and constrain the expansion of the universe~\cite{LIGOScientific:2025jau}. Current approaches, however, rely on hierarchical Bayesian analysis (HBA) methods built on per-event parameter estimation (PE), a workflow that becomes increasingly expensive and noisy as the catalog grows~\cite{Farr:2019rap,Essick:2022ojx,Talbot:2023pex,Heinzel:2025ogf}. As detection rates rise with next-generation observatories such as Einstein Telescope~\cite{Branchesi:2023mws} and Cosmic Explorer~\cite{Reitze:2019iox}, these computational and statistical bottlenecks will be exacerbated.

Population inference starts from a forward model $p_\text{pop}(\theta\mid\Lambda)$ predicting event parameters $\theta$ from population hyperparameters $\Lambda$. Given $N$ observed events, with data $\Data_i$ for event $i$, the population likelihood is~\cite{Mandel:2018mve,Fishbach:2018edt,Vitale:2020aaz}
\begin{equation}\label{eq:pop-likelihood}
    p(\{\Data_i\}_{i=1}^{N}\mid\Lambda) = \prod_{i=1}^{N} \frac{\int p(\Data_i\mid \theta) p_\text{pop}(\theta \mid \Lambda)\,\mathrm{d}\theta}{\int p_\text{det}(\theta)\,p_\text{pop}(\theta\mid \Lambda)\, \mathrm{d}\theta}.
\end{equation}
%which includes integrals for each event that must be evaluated for each choice of hyperparameters.
Here, $p(\Data\mid\theta)$ is the single-event likelihood \cite{LIGOScientific:2025hdt, Talbot:2025vth}. The denominator accounts for selection effects, with $p_\mathrm{det}(\theta)$ the probability that an event with parameters $\theta$ is observed.

The integrals in Eq.~\eqref{eq:pop-likelihood} are typically evaluated using Monte Carlo estimates, introducing uncertainty in the likelihood computation. For the numerator, single-event PE samples computed under a fiducial prior are reweighted to the population prior $p_\mathrm{pop}(\theta\mid\Lambda)$. Meanwhile, the selection function in the denominator is evaluated by importance sampling from a set of simulated events passing a detection threshold \cite{Essick:2025zed, Tiwari:2017ndi}.
% Due to the finite sample sizes, the Monte Carlo log likelihood  variance between hyperparameter points
Due to finite sample sizes, the Monte Carlo variance in the estimate of the log population likelihood % Eq.~\eqref{eq:pop-likelihood}
can scale as badly as $N^2$ \cite{Farr:2019rap,Essick:2022ojx,Talbot:2023pex,Heinzel:2025ogf}.

A separate problem is that standard analyses require evaluating $p_\mathrm{pop}(\theta\mid\Lambda)$, which makes it difficult to compare observations directly to astrophysical simulations without utilizing surrogate models \cite{Barrett:2016edh, Taylor:2018iat, Zevin:2020gbd, Wong:2020ise, Mould:2022ccw, Colloms:2025hib, Plunkett:2025mjr}. In previous work~\cite{Leyde:2023iof}, some of us showed how neural posterior estimation (NPE)~\cite{Papamakarios:2019fms,lueckmann2017flexible,greenberg2019automatic} can be applied to GW population analysis. This removed the need for explicit likelihood evaluations by instead training neural networks for inference based on simulated catalogs (including selection effects). However, our work still relied on PE samples as an intermediate step and required fixing the catalog size in advance of training.

\begin{figure*}[t]
    \begin{tikzpicture}[baseline=(current bounding box.north)]
    \begin{scope}[local bounding box=inference]
        \matrix (flow) [matrix of nodes, nodes={data}, column sep=4mm, row sep=14mm]
        {
            % Row 1: input data
            |[realdata]| $\Data_1$ &
            |[realdata]| $\Data_2$ &
            |[textbox]| $\ldots$   &
            |[realdata]| $\Data_N$ \\
            % |[textbox]| $\phantom{\Data_N}$ \\  % empty spacer for CLS column
        
            % Row 2: embeddings and CLS
            $\embedding_1$ &
            $\embedding_2$ &
            |[textbox]| $\ldots$       &
            $\embedding_N$ \\
            % |[nn, name=clstoken]| {$\embedding_{N+1}$} \\
        };

        \coordinate (midrow) at ($(flow-1-1.south)!0.5!(flow-2-1.north)$);
        \node[nn, wide] (embedding) at (midrow -| flow.center) {\dingo{} embedding $\embedding_i = f(\Data_i)$};
        
        \node[nn, below=of flow, wide] (model) {Transformer encoder};

        \foreach \j in {1,2,4} {%
            \draw[edge] (flow-1-\j.south) -- (embedding.north -| flow-1-\j.south);
            \draw[edge] (embedding.south -| flow-2-\j.north) -- (flow-2-\j.north);
            \draw[edge] (flow-2-\j.south) -- (model.north -| flow-2-\j.south);
        }
        % \draw[edge] (clstoken.south) -- (model.north -| clstoken.south);

        % \node[data, anchor=north] (summary) at ($(model.south -| clstoken.south) + (0,-4mm)$) {Summary token \\ $\embedding_\mathrm{pop}$};
        \node[data, below=of model] (summary) {Summary token \\ $\embedding_\mathrm{pop}$};

        \node[nn, anchor=north] (nf) at ($(summary.south) + (0,-4mm)$) {Normalizing flow};

        \draw[edge] (model.south -| summary.north) -- (summary);
        \draw[edge] (summary) -- (nf);

        \node[data, below=of nf] (hyper) {Hyperparameter samples\\$\Lambda \sim q_\phi(\Lambda \mid \embedding_\mathrm{pop})$};

        \draw[edge] (nf) -- (hyper);

        \begin{pgfonlayer}{background}
            \node[group, fit=(model) (nf), draw=okamber, solid, label={[anchor=south east, inner sep=3pt, font=\footnotesize, text=okamber]south east:{\dingoP}}] (dingopop) {};
        \end{pgfonlayer}

        \node[imgbox, anchor=east, draw=black!60] at ($(hyper.west) + (-2mm,+5mm)$) {\includegraphics[width=3.5cm]{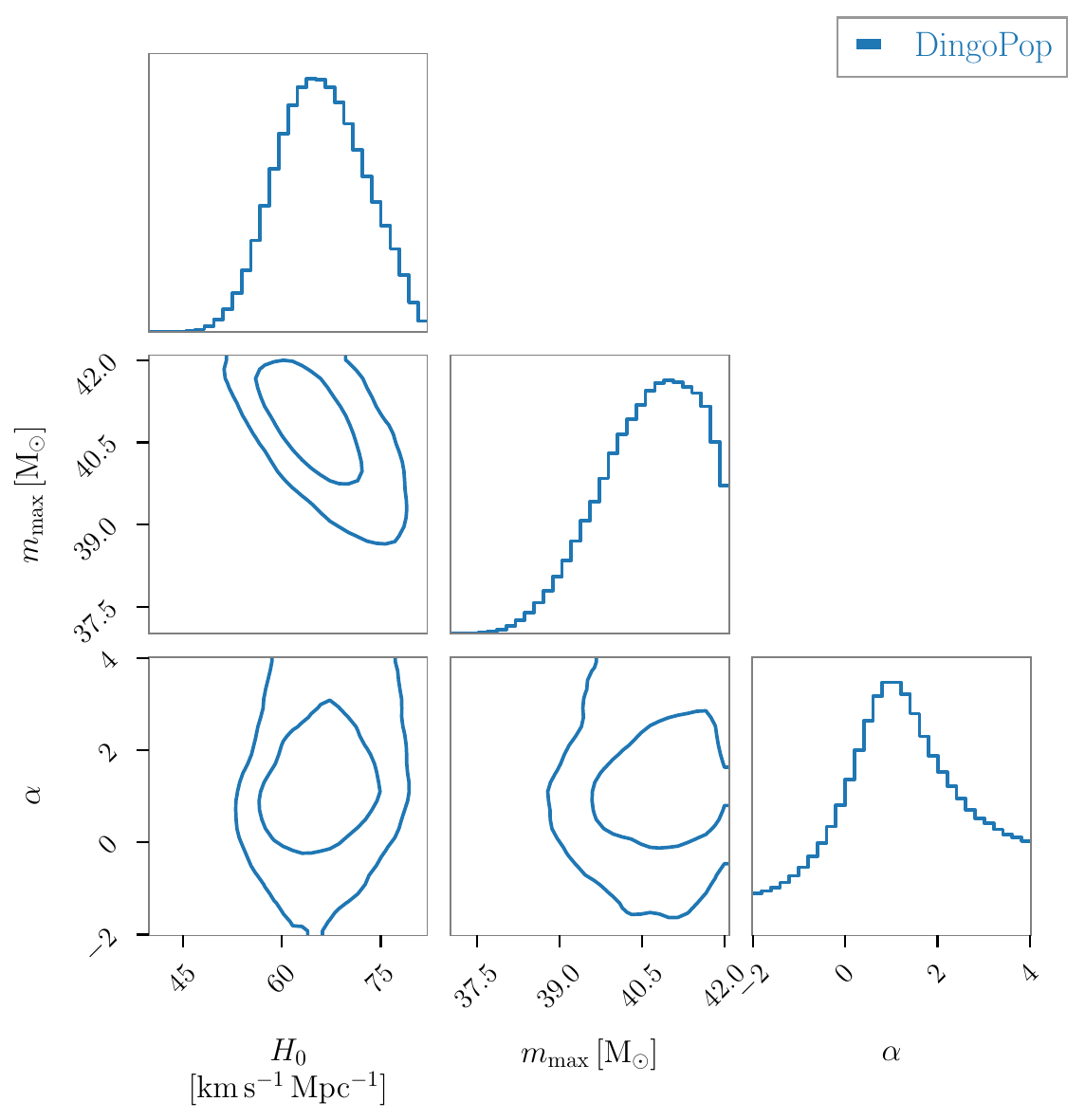}};
    \end{scope}

    \begin{scope}[shift={(inference.north east)}, xshift=4cm, yshift=-5.5mm, name prefix=right-, local bounding box=rightpic]
        % Hyperparameters and population size
        \node[dist] (lambda) {Hyperparameters\\$\Lambda \sim p(\Lambda)$};
        \node[dist, right=of lambda] (N) {Population size\\$N\sim p(N)$};
        \node[inner sep=0, fit=(lambda) (N)] (ingroup) {};

        % Sample events
        \node[dist, below=of ingroup] (theta) {Event parameters\\$\theta \sim \ppop$};
        \node[nn, below=of theta] (pdet) {Detection filter\\$p_\text{det}(\theta)$};
        % \node[decision, below=of pdet] (rejection) {Accept?};
        \node[nn, below=of pdet] (embedding) {Detected embedding\\$\embedding \sim p(\embedding\mid\theta,\text{det})$};

        \draw[edge] (theta) -- (pdet);
        %\draw[edge] (pdet) -- (rejection);
        \draw[edge] (pdet) -- node[right, note]{yes} (embedding);
        \draw[edge] (pdet.west) to[out=160, in=225, looseness=1.0] node[left, note] (no) {no} (theta.south west);

        \node[group, fit=(theta) (pdet) (embedding) (no), label={[xshift=0mm,yshift=4mm]south east:{$\times N$}}] (generation) {};

        \draw[edge] (lambda) -- (generation);
        \draw[edge] (N) -- (generation);

        % Population of events
        \node[data, below=of generation] (events) {Token set $\{\embedding_i\}_{i=1}^N$};
        
        % Population transformer model
        \node[nn, below=of events, wide] (model) {\dingoP model\\ \footnotesize{(transformer + normalizing flow)}};

        % Loss
        \node[data, below=of model, wide] (loss) {Loss (NLL)\\$\mathcal{L}=-\tfrac{1}{B}\sum_{b=1}^B \log q_\phi(\Lambda^b \mid \embedding^b_\mathrm{pop})$};

        % Edges
        \draw[edge] (generation) -- (events);
        \draw[edge] (events) -- (model);
        \draw[edge] (model) -- (loss);

        % Hyperparameter edge
        \draw[edge] ([xshift=-10mm]lambda.south) to[in=120, out=-120] ([xshift=-10mm]lambda.south |- model.north);

        % Indicate batches
        \node[inner sep=0, fit=(lambda) (N) (events)] (batch) {};
        \draw[decorate,decoration={brace}]
        ($(batch.north east)+(2mm,1mm)$) -- ($(batch.south east)+(2mm,-1mm)$)
        node[midway, right, xshift=2mm, note]{\rotatebox{90}{Batch of $B$ populations}};

    \end{scope}

    % Zoom arrows from the group on the left to the detailed model on the right
    \draw[dashed, okamber] (dingopop.north east) -- (right-model.north west);
    \draw[dashed, okamber] (dingopop.south east) -- (right-model.south west);

    \node[font=\bfseries] at ($(flow.north)+(0,6mm)$) {Inference};

    \node[font=\bfseries] at ($(right-rightpic.north)+(0,6mm)$) {Training};
    
    \end{tikzpicture}

  \caption{\dingoP framework, showing inference (left) and training (right). Blue boxes indicate sampling from the population forward model, orange boxes neural networks, cyan boxes network outputs, and purple observed data. To accelerate training, we use auxiliary neural networks that estimate $p_\text{det}(\theta)$ and emulate \emph{detected} embeddings. 
  }
  \label{fig:training-flowchart}
\end{figure*}

In this \emph{Letter}, we introduce \dingoP, an end-to-end simulation-based inference (SBI)~\cite{Cranmer:2019eaq} framework for population analysis that operates directly on GW strain data (Fig.~\ref{fig:training-flowchart}, left). The strain data from individual events are first encoded into compact latent representations using a \dingo embedding network~\cite{Dax:2021tsq}. These embeddings serve as tokens for a transformer encoder \cite{2017arXiv170603762V, Paszke:2019xhz} that accommodates variable catalog sizes. Finally, the transformer output is mapped to a population posterior using a normalizing flow. To train, we simulate populations according to the likelihood \eqref{eq:pop-likelihood} ($\Lambda \to \{\theta_i\} \to \{\mathcal D_i\}$),
% which makes Monte Carlo sampling tractable,
avoiding the expensive likelihood evaluations of HBA.  
% By eliminating single-event PE, \dingoP reduces Monte Carlo noise.
In addition, as an SBI method, \dingoP enables training directly on astrophysical simulations. Transformers have also been applied in general-purpose amortized SBI~\cite{gloeckler2024all}. While transformers have recently been applied to population inference~\cite{Jiang:2025jxt}, \dingoP is the first framework to move beyond PE-based workflows and fixed catalog sizes. We further develop an efficient training pipeline that generates detected catalogs by combining population draws with detection-probability and strain-embedding surrogate networks (Fig.~\ref{fig:training-flowchart}, right).

In our experiments, we train \dingoP on a parametric \plp mass model \cite{Talbot:2018cva}, augmented with cosmological parameters estimated through the mass-spectrum method~\cite{Taylor:2011fs, Farr:2019twy, Mastrogiovanni:2021wsd}. We allow for catalog sizes ranging from  25 to 1000 events. Once trained, inference from strain data takes approximately 1~s and (on simulated data) agrees with standard HBA to within its Monte Carlo uncertainty. %Such fast inference opens new possibilities for systematics studies and forecasting for future detectors. % that were computationally infeasible until now.
As an application, we explore the dependence of the spectral-siren Hubble constant uncertainty on catalog size. % and composition.

% other references: Mancarella:2021ecn, Leyde:2022orh, Ezquiaga:2022zkx, Pierra:2023deu, Bom:2024afj, Farah:2024xub, MaganaHernandez:2024uty, Mali:2024wpq, Agarwal:2024hld, Li:2024rmi, Tong:2025xvd, LIGOScientific:2025jau, MaganaHernandez:2025cnu

\sec{Method}A key obstacle in treating individual GW events as tokens stems from the dimension and complexity of strain data. GW data for a black hole binary has dimensionality $\sim\!10^5$, consisting of a signal in multiple detectors (potentially with precession and higher modes) deeply embedded in detector noise (assumed stationary Gaussian). Interpreting each event in terms of the source parameters is thus already a challenging task for a neural network. We therefore leverage pre-trained single-event \dingo models to compress events into low-dimensional embeddings. A \dingo model consists of an embedding network that encodes raw strain data and power spectral densities $\Data$ into a low-dimensional summary $\embedding = f(\Data)$, followed by a normalizing flow conditioned on $\embedding$ for PE. Here, we use just the encoder part of a trained \dingo model, obtaining a low-dimensional summary that contains the information needed to efficiently construct the full posterior. To simplify the representation, we use a specialized \dingo model that infers only the parameters relevant for our population model (masses and luminosity distance), with a 32-dimensional embedding.

We pass each embedding $\embedding_i$ for event $i$ through a residual network to produce tokens. The population posterior does not depend on event orderings, so (in contrast to other applications) we do not include any positional information with the tokens; the naturally permutation-invariant transformer architecture then preserves this invariance~\cite{zaheer2017deep,lee2019set}. To the event token sequence, we prepend a learnable summary token~\citep{devlin:2019, darcet_registers:2024}, which aggregates information from the event tokens through self-attention layers. Denoting the tokenizer, summary token, and transformer encoder jointly by $T_\psi$, we obtain the output summary $\embedding_\text{pop} = T_\psi(\{\embedding_i\}_{i=1}^N)$. We take this as context for a normalizing flow, which learns an approximation $q_\phi(\Lambda\mid\embedding_\text{pop})$ to the population posterior. The architecture is illustrated in Fig.~\ref{fig:training-flowchart}. Self-attention scales quadratically with catalog size; however, for our studies, training is still bottlenecked by data generation. For larger catalogs, alternatives to full self-attention may be needed, e.g., sparse or linear attention~\cite{katharopoulos2020transformers,Kitaev2020Reformer,child2019generatinglongsequencessparse,beltagy2020longformer}, or decoupled set-encoding approaches~\cite{wehenkel2026justtakestwoscaling}.%, or one could explicitly suppress self-attention between events, resulting in linear scaling.

We train \dingoP using the negative log-likelihood loss,
\begin{equation}\label{eq:loss}
    \mathcal{L} = \mathbb{E}_{p(\Lambda)p(N)p(\{\Data_i\}_{i=1}^N\mid\Lambda)}\left[-\log q_\phi\left(\Lambda \mid T_\psi(\{f(\Data_i)\}_{i=1}^N)\right)\right],
\end{equation}
but with several optimizations discussed below. Here, the learnable flow and transformer parameters $\phi$ and $\psi$ are jointly optimized with the embedding $f$ fixed, $p(\Lambda)$ is the hyperparameter prior, and $p(N)$ is a uniform distribution over catalog sizes. Given $N$ and $\Lambda$, we generate a source population
% based on the likelihood \eqref{eq:pop-likelihood},
including selection effects to obtain detected events. For each population seen in training, we mask events beyond a random size $N$; this trains the network to interpret variable catalog sizes. At inference time, we pass the actual observed events (for any $N$).

Because of selection effects, the data generative process involves producing large numbers of events that are never detected, which is computationally wasteful. Moreover, our networks (with $\sim\!10^8$ trainable parameters) are prone to overfitting if populations are re-used excessively in training. Thus, we developed an efficient approach to accelerate data production, enabling on-the-fly generation during training: two auxiliary networks that allow us to avoid generating unobserved events and to produce data directly in the embedding space (Fig.~\ref{fig:training-flowchart}, right). First, we train a network to estimate the detection probability $p_\text{det}(\theta)$, marginalized over event parameters not directly predicted by the population model (e.g., everything but $m_{1,\text{d}}, m_{2,\text{d}}, d_L$ in our case) and over noise (see Supplemental Material). % \cite{Talbot:2020oeu, Gerosa:2020pgy, Callister:2024qyq, Lorenzo-Medina:2024opt};
After sampling event parameters, we use rejection sampling to decide whether a given event is observed. Second, we train a conditional normalizing flow as an emulator for detected embeddings, $p(\embedding\mid \theta, \text{det})$, with the same marginalization; this allows us to directly sample an embedding $\embedding_i$ given that event $i$ is observed.

\sec{Results}For our population model, we combine the \plp mass distribution with a flat $\Lambda$CDM cosmology (fixed $\Omega_\text{m}=0.3$, variable $H_0$). 
Primary masses follow a power law $\propto\msone^{-\alpha}$ between the minimum and maximum masses, $m_\text{min}$ and $m_\text{max}$, with a fraction $\lambda_{\rm g}$ falling into a truncated Gaussian with mode $\mug$ and width $\sigmag$; secondary masses follow a power law $\propto\mstwo^{\beta}$ between $m_\text{min}$ and $\msone$; both mass distributions are tapered from $m_\mathrm{min}$ over a mass interval of width $\delta_{\rm m}$ (see App.~A of \cite{LIGOScientific:2021aug}).
Redshifts $z$ are drawn according to a constant merger rate over comoving volume and source-frame time; the luminosity distance $d_\text{L}$ follows from $z$ and the cosmological parameters. All other source properties follow standard distributions. 
GW signals are generated using the \texttt{IMRPhenomXPHM} precessing, higher-mode waveform model~\cite{Pratten:2020ceb, Ramos-Buades:2023ehm}, to which we add stationary Gaussian noise. We assume a two-detector configuration of LIGO Hanford and Livingston at O3 sensitivity \cite{LIGOScientific:2014pky,aLIGO:2020wna,Tse:2019wcy, KAGRA:2021vkt}.  We use a signal-to-noise ratio of 12 as our detection threshold. 
In total, the population model is described by nine hyperparameters: $H_0$, $\mmin$, $\mmax$, $\alpha$, $\beta$, $\lambda_{\rm g}$, $\mug$, $\sigmag$, and $\delta_{\rm m}$. Priors are constrained by the single-event \dingo model, which covers detector-frame component masses up to 100~$M_\odot$, limiting $m_\text{max}$ to 42~$M_\odot$, see the Supplemental Material.

\begin{figure}[t]
    \centering
    \includegraphics[width=1.0\linewidth]{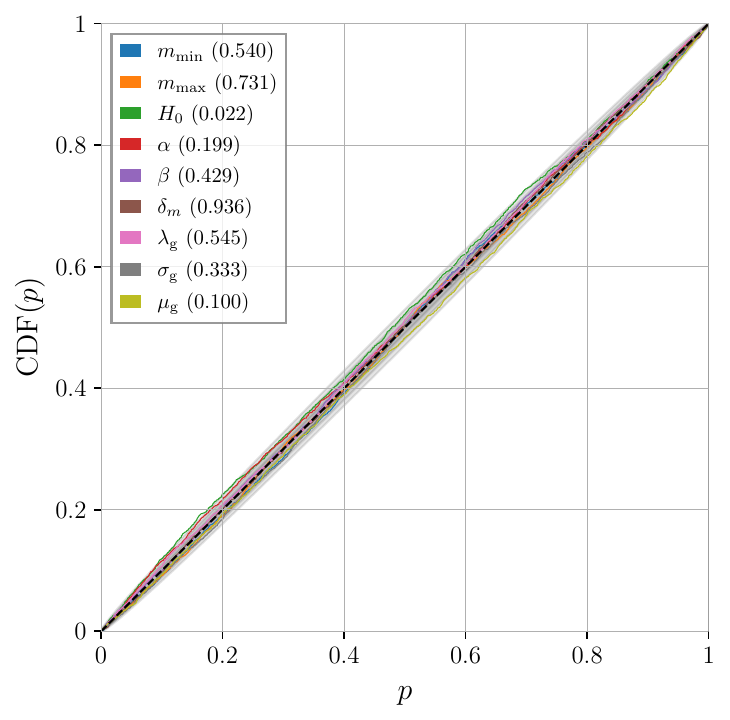}
    \caption{P--P plot for 2500 simulated catalogs with random sizes $N\sim \mathcal{U}(25,1000)$. 
    Gray bands show 1-, 2-, and 3-$\sigma$ intervals under perfect calibration; Kolmogorov--Smirnov $p$-values are listed in the legend (combined $p$-value: 0.20). 
    }
    \label{fig: pp plot}
\end{figure}

The \dingoP network uses a 10-layer transformer encoder with embedding dimension 1024, followed by a 14-step neural spline flow \cite{Durkan:2019nsq}, with $1.5\times 10^8$ trainable parameters in total.
During training, the network sees \totalpopulationsSci{} unique populations (\totalsamplesSci{} total GW events), divided into batches of \batchsizesimple{}. Training takes 11~days on an NVIDIA A100 GPU (see the Supplemental Material).

We validate \dingoP on simulated catalogs, both with probability--probability (P--P) plots to check model calibration and with comparisons against standard HBA. The P--P plot in Fig.~\ref{fig: pp plot} is based on $2500$ simulated catalogs, each with randomly chosen $N$. We draw 5000 posterior samples per catalog. For each hyperparameter, we compute the percentile score of the true value within its one-dimensional marginal, and we plot the cumulative distributions of these scores across all catalogs. For well-calibrated posteriors, the percentiles should be uniformly distributed---indeed, Fig.~\ref{fig: pp plot} shows that the distributions are consistent with the diagonal (combined $p$-value of 0.20). By randomly sampling the catalog sizes, this test is sensitive to the full range of $N$ used in training. 
To our knowledge, a calibration test of this scale is unprecedented; even producing the requisite per-event PE samples by conventional means would require $\sim\!10^7$~CPU-hours---compared to less than an hour of inference time for \dingoP.%, assuming (optimistically) 10~CPU-hours per event. This can be contrasted with the \dingoP that takes a $\mathcal{O}(10)$~hours for the inference of 2500 populations. 

%To validate \dingoP, we analyze $2500$ independent simulated catalogs of $N=1000$ detected events each and construct a P-P plot (Fig.~\ref{fig: pp plot}, left). The hyperparameters are well calibrated; for $H_0$, $\alpha$, $\mug$, and $\sigma_g$, small but statistically detectable deviations correspond to biases of at most \srg{XX} percentage points. The right panel of Fig.~\ref{fig: pp plot} shows how these deviations vary with catalog size. An injection study of this scale is, to our knowledge, unprecedented and is only feasible due to the amortized nature of the inference and by bypassing single-event PE. \srg{It would be better to fully cover the range of $N$ with the PP plot. Can we randomly sample $N$ as well when making this? We could also use just 1000 injections, and we should report the combined $p$-value.}

\begin{table}[t]
\caption{Comparison of \dingoP and conventional HBA applied to two example populations, showing medians and 90\% credible regions for each hyperparameter. Each catalog consists of 500 events.}
\renewcommand{\arraystretch}{1.3}
\begin{ruledtabular}
\begin{tabular}{l c c c @{\hspace{1.5mm}\vrule width 0.3pt\hspace{1.5mm}} c c c}
  & \multicolumn{3}{c}{Population 1} & \multicolumn{3}{c}{Population 2} \\
$\Lambda$ & Truth & SBI & HBA & Truth & SBI & HBA \\
\hline
$m_{\rm min}$ & 18.1 & $18.7^{+0.9}_{-0.9}$ & $18.5^{+0.8}_{-0.8}$ & 21.6 & $21.1^{+1.0}_{-1.2}$ & $21.3^{+0.8}_{-0.9}$ \\
$m_{\rm max}$ & 40.8 & $40.7^{+1.1}_{-1.7}$ & $40.8^{+1.0}_{-1.1}$ & 38.4 & $40.4^{+1.4}_{-2.1}$ & $40.9^{+1.0}_{-1.8}$ \\
$H_0$ & 64.7 & $65.8^{+10.9}_{-9.9}$ & $65.6^{+10.0}_{-9.7}$ & 72.3 & $68.1^{+9.9}_{-10.9}$ & $67.8^{+9.6}_{-8.6}$ \\
$\alpha$ & 1.1 & $1.2^{+2.4}_{-2.4}$ & $1.1^{+2.4}_{-2.3}$ & $-0.4$ & $2.3^{+1.6}_{-3.6}$ & $2.3^{+1.5}_{-3.5}$ \\
$\beta$ & $-1.1$ & $-0.5^{+2.5}_{-1.3}$ & $-0.8^{+1.3}_{-1.0}$ & 2.3 & $1.5^{+2.2}_{-2.9}$ & $1.6^{+2.1}_{-2.6}$ \\
$\delta_m$ & 4.7 & $3.5^{+2.1}_{-1.3}$ & $4.0^{+1.7}_{-1.8}$ & 3.8 & $4.3^{+1.5}_{-2.0}$ & $4.3^{+1.5}_{-2.0}$ \\
$\lambda_{\rm g}$ & 0.0 & $0.3^{+0.5}_{-0.3}$ & $0.3^{+0.4}_{-0.3}$ & 0.7 & $0.7^{+0.3}_{-0.5}$ & $0.8^{+0.2}_{-0.5}$ \\
$\sigma_{\rm g}$ & 6.0 & $7.5^{+2.3}_{-5.0}$ & $7.7^{+2.1}_{-5.0}$ & 7.6 & $5.6^{+3.0}_{-3.0}$ & $5.9^{+2.4}_{-2.5}$ \\
$\mu_{\rm g}$ & 25.0 & $28.7^{+5.4}_{-7.6}$ & $28.4^{+5.9}_{-7.4}$ & 21.1 & $25.2^{+3.7}_{-4.6}$ & $25.1^{+3.3}_{-4.4}$ \\
\end{tabular}
\end{ruledtabular}
\label{tab:HBAcomparison}
\end{table}

\begin{figure}[t!]
\includegraphics[width=\linewidth]{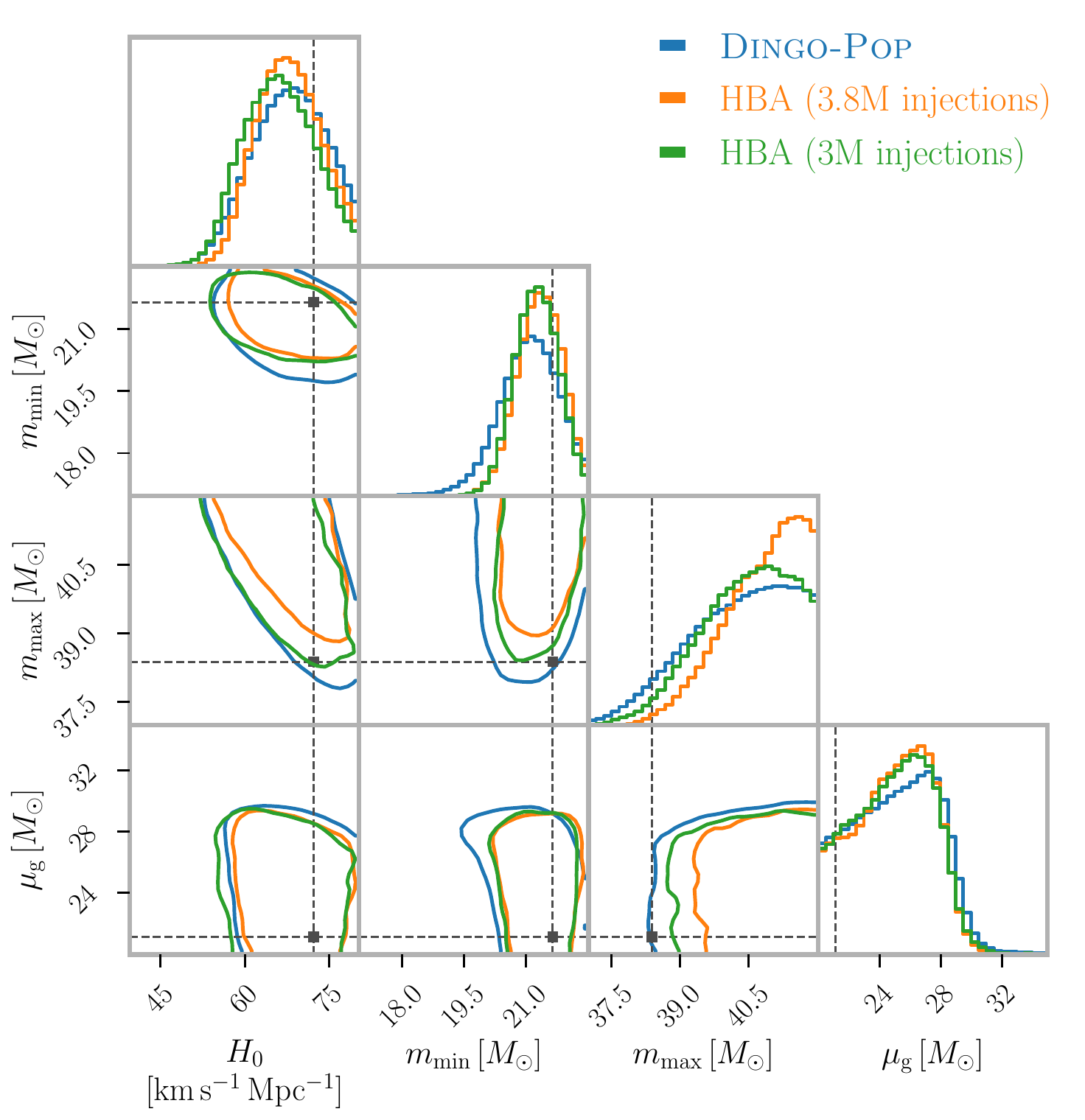}

\vspace{3mm}

\hspace*{-5.5mm}
\includegraphics[width=1.02\linewidth]{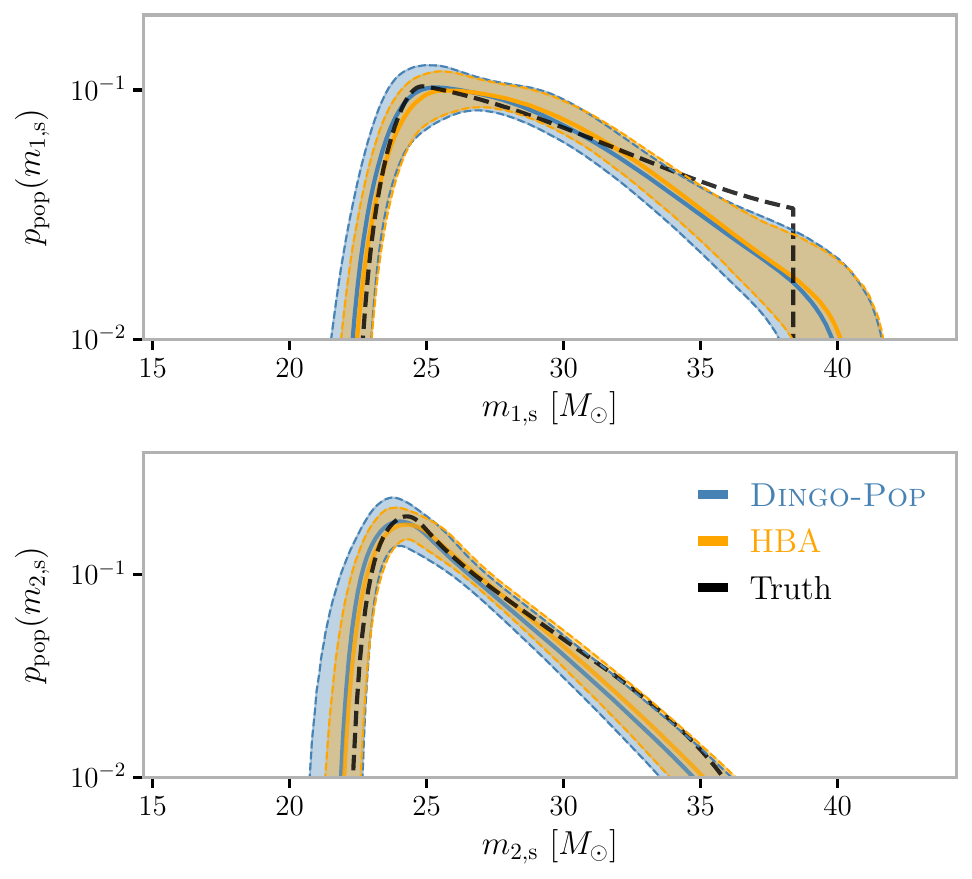}
\caption{Hyperparameter posterior (top) and inferred mass spectra (bottom) from Population 2. We compare \dingoP (blue) with two HBA analyses (orange: $3.8\times 10^6$ injections; green: $3\times 10^6$; the mass spectra panel only shows the higher-injection HBA). 
Contours and shaded bands give 90\% credible regions; dashed black lines mark the true population. Full corner plots for both populations are in the Supplemental Material.
    }
    \label{fig:corner}
\end{figure}

Passing a P--P test does not guarantee accurate posteriors. 
Thus, we also compare \dingoP posteriors directly against standard HBA. We analyze two example catalogs of 500 events each. For the conventional analysis, we generate single-event PE samples using a mix of \dingo and \textsc{Bilby} \cite{2019ApJS..241...27A}, and use \icarogw{}~\cite{Mastrogiovanni:2021wsd, Mastrogiovanni:2023emh} for the population analysis. 
Because we applied our selection cut based on true parameters and data (through the matched-filter SNR using the true template) in training \dingoP, we apply a corresponding cut to PE samples in the HBA likelihood for consistency~\cite{Essick:2023upv}.

The median and 90\% credible regions agree across both populations (Tab.~\ref{tab:HBAcomparison}). We further provide a partial corner plot and reconstructed mass spectra for Population 2 in Fig.~\ref{fig:corner}. Some marginals show mild tension between \dingoP and HBA. To assess whether the HBA analysis has fully converged, we perform this analysis with two different injection sets (of sizes $3\times 10^6$ and $3.8\times 10^6$). We find that the deviation between HBA analyses is of similar order to the discrepancy with \dingoP, indicating that Monte Carlo uncertainty in the selection-function estimate is a dominant source of error; the \dingoP distribution is also slightly broader. Reaching this level of agreement required several iterations of the HBA analysis to identify and reconcile subtle effects (such as the selection-cut consistency noted above), with \dingoP serving as a rapid independent check. The \dingoP mass spectra are also very slightly broader than HBA.

\begin{figure}[t]
    \centering
    \includegraphics[width=1.0\linewidth]{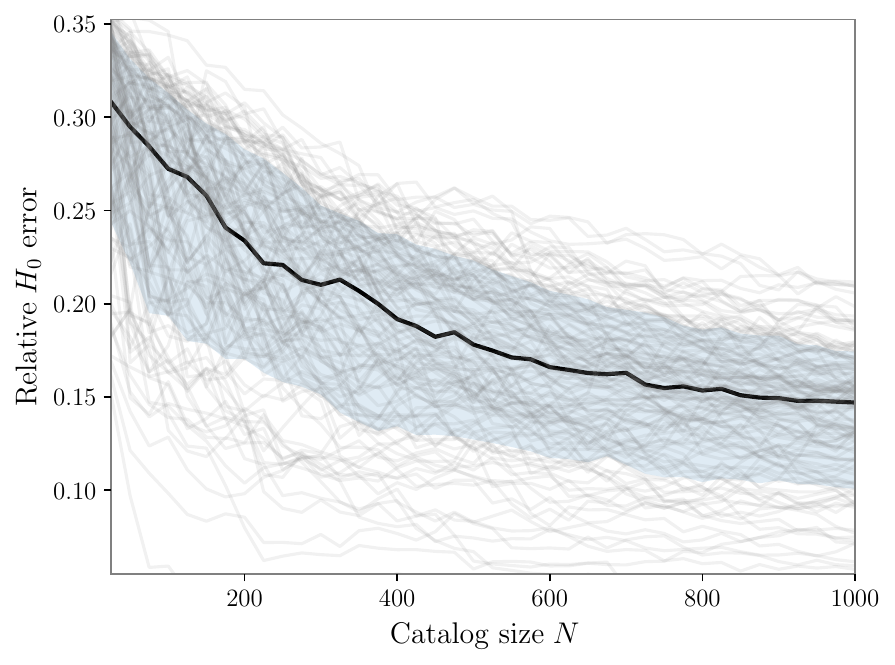}
    % \vspace{1.2cm}
    \caption{Relative $H_0$ uncertainty (2-$\sigma$ width divided by median) versus catalog size for 128 simulated populations. Gray: individual populations as events are added; black: median; blue band: 1-$\sigma$ scatter across populations. Plots for the other hyperparameters are in the Supplemental Material.
    }
    \label{fig:examples_applications}
\end{figure}

Finally, as an application that leverages the speed of \dingoP, we perform an injection study to explore the dependence of spectral-siren $H_0$ constraints on catalog size. We simulate 128 populations with hyperparameters drawn from the prior and, for each, we compute the \dingoP posterior as events are added incrementally. Fig.~\ref{fig:examples_applications} shows the relative $H_0$ uncertainty as a function of $N$. The uncertainty decreases with $N$, though with significant population-to-population scatter reflecting the specific hyperparameters and catalog realizations. The relative $H_0$ uncertainty reaches $\sim\!23\%$ at 200 events and $\sim\!15\%$ at 1000 events, with a $N^{-0.28}$ scaling between 500 and 1000 events. These numbers are sensitive to our narrow $m_{\text{max}}$ prior, imposed by the underlying \dingo model.

\sec{Conclusions}We presented \dingoP, a transformer-based framework for end-to-end GW population inference directly from strain data. It circumvents per-event parameter estimation (and the associated biases~\cite{Farr:2019rap,Essick:2022ojx,Talbot:2023pex,Heinzel:2025ogf}), allows for variable catalog sizes, and gives inference in just 1\,s. We trained a network for catalogs of up to $10^3$ events and validated it using P--P plots and direct comparisons against standard HBA. 
%As an SBI method, \dingoP could also be trained directly on astrophysical population-synthesis simulators, beyond the parametric model considered here. 

There remain several steps before applying our framework to real data. First, the detection threshold for the population simulator must be made to match that of the LVK detectors, e.g., training the $p_\mathrm{det}(\theta)$ network based on LVK pipeline injections~\cite{Talbot:2020oeu, Gerosa:2020pgy, Callister:2024qyq, Lorenzo-Medina:2024opt}. Second, the underlying \dingo embedding network must be extended to cover the full range of observed events, detector configurations, and data conditioning settings. Recent transformer-based models for flexible single-event inference provide a roadmap for suitable embeddings~\cite{Kofler:2025dux}.

A further challenge is model misspecification, which is likely to be more severe at the population level than for single events (where general relativity is very well understood). For conventional HBA, one will still infer the ``correct'' hyperparameter posterior for the incorrect population model, whereas NPE can fail due to observations being out of the training distribution.
We demonstrated basic robustness~\cite{cannon2022investigating,schmitt2023detecting,wehenkel2024addressing} in the Supplemental Material, but additional mitigation strategies may be necessary in practice, such as training across a range of population models to better cover plausible catalog realizations.

Beyond the setting of conditionally independent events explored here, the transformer architecture of \dingoP can capture correlations between events. Natural extensions include strongly-lensed event pairs sharing source parameters, or joint analyses with galaxy catalogs or electromagnetic counterparts to account for cross-correlations with GW data. These problems lie outside the scope of methods that aggregate per-event posteriors, such as compositional score matching~\cite{geffner2023compositional}.

The speed of \dingoP enables real-time population updates during future observing runs~\cite{Wolfe:2026dcq},
and forecasting and systematics studies at scales out of reach for conventional methods.
As GW catalogs grow to $O(10^{5-6})$ events with next-generation detectors, the approach developed here provides a foundation for fully realizing their scientific potential.

\medskip
\sec{Acknowledgments}We thank N. Gupte, C. Talbot, A. Toubiana, and M. Williams for helpful discussions. 
K.L. is supported by ERC Starting Grant SHADE (grant no.~StG 949572). 
S.R.G. is supported by a UKRI Future Leaders Fellowship (grant number MR/Y018060/1).
M.M. is supported by a Royal Commission for the Exhibition of 1851 research fellowship.
Numerical computations were (in part) carried out on the \texttt{Sciama} High Performance Computing (HPC) cluster, which is supported by the Institute of Cosmology and Gravitation (ICG), the South-East Physics Network (SEPNet) and the University of Portsmouth.
The computations reported in this paper were (in part) performed using resources made available by the Flatiron Institute.
The Center for Computational Astrophysics at the Flatiron Institute is supported by the Simons Foundation.
This material is based upon work supported by NSF's LIGO Laboratory which is a major facility fully funded by the National Science Foundation.

\bibliography{references, reference-software}

\clearpage
\begin{center}
  \large
  \textbf{Supplemental Material}
\end{center}

\section{Prior distributions}
\label{app: properties training dataset}

The \dingoP framework involves two levels of prior distributions: (1) the single-event priors used to train the underlying \dingo embedding network (Tab.~\ref{tab: summary networks single-event prior}) and (2) the population hyperparameter priors used to train \dingoP itself (Tab.~\ref{tab: summary networks prior}). Since the \dingo model is trained only on the parameter ranges in Tab.~\ref{tab: summary networks single-event prior}, these ranges constrain what population models \dingoP can study.

\begin{table}[th]
\caption{Summary of priors used to  generate training sets for the single-event \textsc{Dingo} models, from which only the embedding networks are used in \dingoP. We use the shorthand notation $\sin(0,\pi)$ for a sine distribution between 0 and $\pi$,
%(and similarly for $\cos$). Finally
$\text{ct}$ for a constraint (chirp mass and mass ratio prior simply act as constraints on the component masses),
and $\mathcal{U}$ for a uniform distribution. 
}
\renewcommand{\arraystretch}{1.2}
\centering
\begin{ruledtabular}
\begin{tabular}{lc}
\textbf{Parameter} & \textbf{Prior} \\
\hline
\multicolumn{2}{c}{\textit{Intrinsic }} \\
\hline
Primary mass ($\mdone$) $[\msun]$ & $\mathcal{U}$(10, 100) \\
Secondary mass ($\mdtwo$) $[\msun]$ & $\mathcal{U}$(10, 100) \\
Chirp mass ($\mathcal{M}$) $[\msun]$ & $\text{ct}(15, 120)$ \\
Mass ratio ($q$) & $\text{ct}(0.1, 1.0)$ \\
Orbital phase ($\phi$) [rad] & $\mathcal{U}(0,2\pi)$ \\
Inclination angle ($\theta_{JN}$) [rad] & $\sin(0,\pi)$ \\
Primary spin magnitude ($a_1$) & $\mathcal{U}(0.0, 0.99)$ \\
Secondary spin magnitude ($a_2$) & $\mathcal{U}(0.0, 0.99)$ \\
Primary spin--orbit tilt ($\theta_1$) [rad] & $\sin(0,\pi)$ \\
Secondary spin--orbit tilt ($\theta_2$) [rad] & $\sin(0,\pi)$ \\
Primary--secondary azimuth ($\phi_{12}$) [rad] & $\mathcal{U}(0,2\pi)$ \\
Total--orbital azimuth ($\phi_{JL}$) [rad] & $\mathcal{U}(0,2\pi)$ \\
\hline
\multicolumn{2}{c}{\textit{Extrinsic}} \\
\hline
Right ascension ($\alpha$) [rad] & $\mathcal{U}(0,2\pi)$ \\
Declination ($\delta$) [rad] & $\cos(-\pi/2,\pi/2)$ \\
Geocentric time ($t_g$) [s] & $\mathcal{U}(-0.10, 0.10)$ \\
Polarization angle ($\psi$) [rad] & $\mathcal{U}(0,\pi)$ \\
Luminosity distance ($d_\mathrm{L}$) [Mpc] & $\mathcal{U}(100, 8000)$ \\
\end{tabular}
\end{ruledtabular}
 \label{tab: summary networks single-event prior}
\end{table}

For the population model, we use the \plp mass model \cite{Talbot:2018cva, LIGOScientific:2021aug} but limit both components to the maximum mass $m_{\rm max}$ rather than the Gaussian peak having a separate limit of $\mug + 5 \sigmag$ as in \texttt{icarogw}; otherwise, the mass range for the \dingo model would be exceeded (cf.~Tab.~\ref{tab: summary networks single-event prior}).
Recall that detector-frame ($\md$) and source-frame mass ($\ms$) are related through $\md = (1 + z)\ms$.
Thus, for a given upper limit on the detector-frame mass and a maximum luminosity distance, this implies a maximum source-frame mass. 
To relate the luminosity distance to a redshift, cosmological parameters are needed, and the redshift is maximized for the largest $H_0$ allowed under the prior. 
In our case, for a luminosity distance of 8000~Mpc and a maximum $H_0 = 80~\hu$, this gives a maximum redshift of 1.3, translating into a maximum source-frame mass of $\sim\!43~\msun$.

The hyperparameter priors are summarized in Tab.~\ref{tab: summary networks prior}. For source properties other than masses and distance, we use the same fixed distributions as in Table~\ref{tab: summary networks single-event prior}. For distance, we instead use the  population distribution corresponding to a merger rate that is constant over comoving volume and source-frame time.

\begin{table}[th]
\caption{Summary of distributions for the hyperparameters of the population model used to generate training data for \textsc{Dingo-Pop}, which also corresponds to the Bayesian prior for the trained posterior. The uniform prior is denoted as $\mathcal{U}$.}
\renewcommand{\arraystretch}{1.2}
\centering
\setlength{\tabcolsep}{20pt}
\begin{ruledtabular}
\begin{tabular}{cc}
\textbf{Hyperparameter} & \textbf{Prior} \\
 \hline
 $H_0$ $[\mathrm{km\,s^{-1}\,Mpc^{-1}}]$ & $\mathcal{U}(40,80)$ \\
 $\mmin$ $[\msun]$ & $\mathcal{U}(17.5,22.5)$ \\
 $\mmax$ $[\msun]$ & $\mathcal{U}(37,42)$ \\
 $\alpha$  & $\mathcal{U}(-2,4)$ \\
 $\beta$  & $\mathcal{U}(-2,4)$ \\
 $\delta_{\rm m}$ $[\msun]$ & $\mathcal{U}(2,6)$ \\
 $\lambda_{\mathrm{g}}$ & $\mathcal{U}(0,1)$ \\
 $\mu_{\rm g}$ $[\msun]$ & $\mathcal{U}(20,35)$ \\
 $\sigma_{\rm g}$ $[\msun]$ & $\mathcal{U}(1,10)$ \\
\end{tabular}
\end{ruledtabular}
\label{tab: summary networks prior}
\end{table}

\section{\textsc{Dingo-Pop} architecture}
\label{app: neural network architecture}

\begin{table*}[t]
    \caption{Summary of the \textsc{Dingo-Pop} architecture and auxiliary networks.}
    \centering
    \setlength{\tabcolsep}{10pt}
    \renewcommand{\arraystretch}{1.3}
    \begin{tabular}{l l l}
        \hline\hline
        \textbf{Component} & \textbf{Property} & \textbf{Value} \\
        \hline
        \multicolumn{3}{l}{\textit{\textsc{Dingo-Pop} main network}} \\
        \hline
        Tokenizer
            & Type              & Fully connected residual network \\
            & Blocks            & 5 \\
            & Hidden dimension  & 1024 \\
            & Normalization     & Layer normalization (no batch normalization) \\
            & Positional encoding & None (permutation invariant) \\
        \hline
        Transformer encoder
            & Layers            & 10 \\
            & Embedding dimension & 1024 \\
            & Self-attention heads & 8 \\
            & Feedforward hidden dimension & 1024 \\
            & Pooling           & \texttt{cls}-token \\
        \hline
        Final feedforward network
            & Type              & Fully connected residual network \\
            & Blocks            & 5 \\
            & Hidden dimension  & 1024 \\
            & Output dimension  & 512 \\
        \hline
        Normalizing flow
            & Type              & Neural spline flow (NSF) \\
            & Flow steps        & 14 \\
            & Spline bins       & 8 (rational quadratic) \\
            & Conditioner network & Residual network, 5 blocks, 512 units \\
            & Dimensionality & 9 (hyperparameters) \\
        \hline
        \multicolumn{3}{l}{\textit{Auxiliary networks}} \\
        \hline
        Embedding emulator
            & Type              & Neural spline flow (NSF) \\
            & Flow steps    & 28 \\
            & Spline bins & 8 (rational quadratic) \\
            & Conditioner network & Residual network, 5 blocks, 512 units \\
            & Conditioning inputs & $m_{1,\mathrm{d}},\, m_{2,\mathrm{d}},\, d_L$ \\
            & Output dimensions & 32 (embedding dimension) \\
        \hline
        Detection probability
            & Type              & Fully connected residual network \\
            & Blocks & 5 \\
            & Hidden dimension & 128 \\
            & Detection criterion & Matched-filter SNR $\geq 12$ \\
            & Conditioning inputs & $m_{1,\mathrm{d}},\, m_{2,\mathrm{d}},\, d_L$ \\
            & Output dimension & 1 (detection probability) \\
        \hline\hline
    \end{tabular}
    \label{tab: architecture_summary}
\end{table*}

In this work, we adopt a transformer-based architecture \cite{2017arXiv170603762V} for population NPE. The model has four components: a tokenizer, a transformer encoder, a final feedforward network, and a normalizing flow that estimates the hyperparameter posterior. The architectures of \dingoP and its two auxiliary networks are detailed in Tab.~\ref{tab: architecture_summary}. The neural networks are implemented in \textsc{PyTorch} \cite{Paszke:2019xhz}, with layer normalization (rather than batch normalization) throughout.

First, the tokenizer transforms the \dingo embeddings for each event into a high-dimensional embedding space, resulting in a sequence of $N$ tokens.

These are then processed by a transformer, which transforms the sequence through 10 successive self-attention layers. In each such layer, token representations are updated through a weighted sum over the sequence, with weights determined by learned pairwise compatibilities. This embeds the tokens in the global context, allowing for interactions across the sequence. 
We further append a learnable, randomly initialized \texttt{cls}-token~\citep{devlin:2019, darcet_registers:2024} to the sequence, which then interacts with the tokens throughout the transformer layers, aggregating information from the full sequence. 
Transformers are well suited to population inference for two reasons: they naturally handle variable-length inputs, enabling a single network to amortize across catalogs of different sizes; and in the absence of positional encodings, self-attention is permutation-equivariant, which matches the permutation invariance of the population posterior under reordering of events.

From the transformer output, we only extract the \texttt{cls}-token, which acts as a global summary of the data. This is then processed with a final feedforward neural network. 

The output of this network then conditions a neural spline flow (NSF) \cite{Durkan:2019nsq}, which estimates the population posterior over the nine hyperparameters.

\section{Training}

We train \dingoP using the \texttt{AdamW} optimizer with initial learning rate $8.0 \times 10^{-5}$ and weight decay $0.01$. The learning rate follows cosine annealing from this value to zero over 800 epochs. Training takes $\sim\!11$~days on an A100 GPU, with 50,000 populations per epoch (batch size 128). Inference takes 1.1~s per 5,000 samples.

We use two auxiliary networks (the detection probability estimator and the embedding emulator) to accelerate training, replacing waveform generation and embedding computation with fast emulation.

\begin{figure*}[!t]
    \centering
    \includegraphics[width=0.97\linewidth]{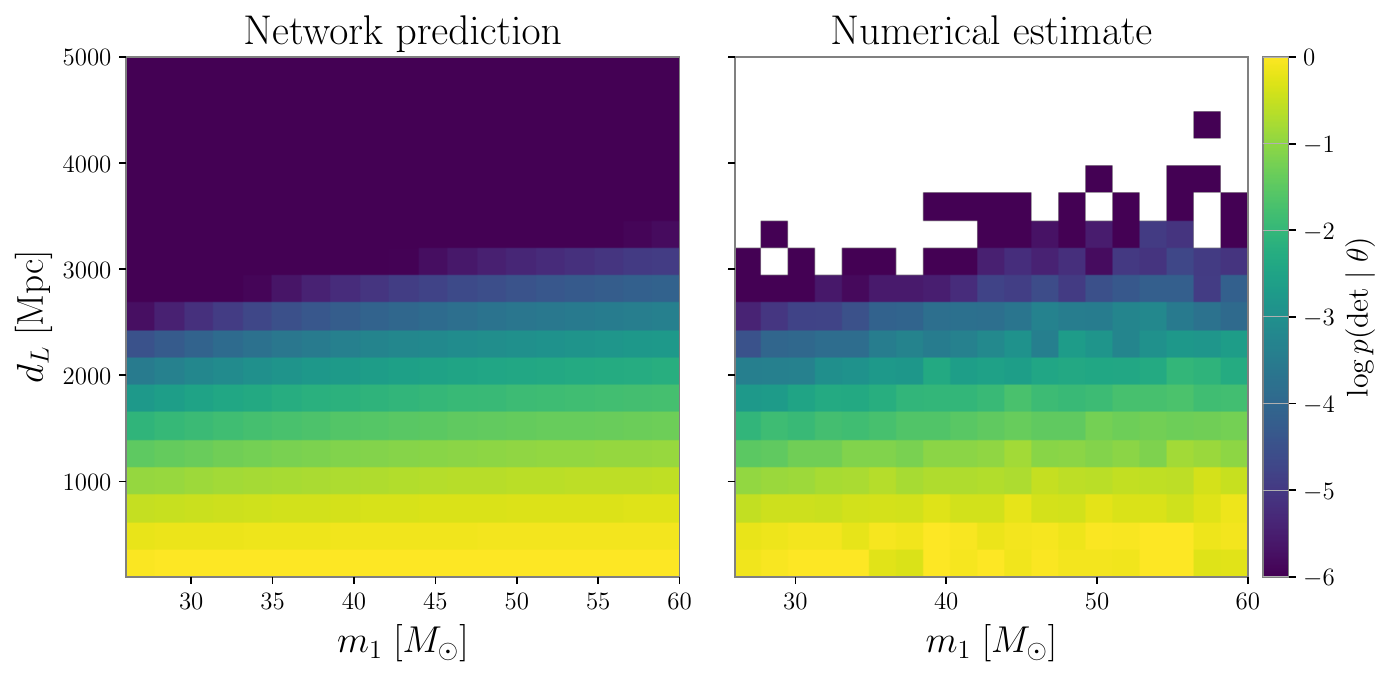}
    \caption{Log detection probability as a function of detector-frame mass and luminosity distance, for secondary mass of $m_{2, {\rm d}} = 26~\msun$.
    The left panel estimates the detection probability from the detection probability estimator, while the right panel relies on $2.5\times 10^6 $ injections.
    Note the non-negligible scatter of the estimated detection probability through samples due to Poisson noise. Also clearly apparent is the detection horizon where GW detectability decreases by a factor of $O(e^{6})$.
    }
    \label{fig: pdet example}
\end{figure*}

\subsection{Detection probability estimator}

The detection probability for an event with source parameters $\theta$ is
\begin{equation}
    \p(\det\mid\theta)
     = \int
     \p(\det\mid\Data)
     \p(\Data \mid \theta) 
     \mathrm{d}\Data
     \,,
\end{equation}
which marginalizes over all possible noise realizations to which signals are added.
For real data, the detection function $p(\mathrm{det}\mid\mathcal{D})$ is a deterministic function of data $\mathcal{D}$ alone; using an SNR threshold computed by filtering data against signal templates for the exact true source properties is not realistic and instead implies a detection function $p(\mathrm{det}\mid\mathcal{D},\theta)$ \cite{Essick:2023upv}.
% , but several recent population studies suggest that any bias from this assumption does not impact realistic population analyses \cite{Heinzel:2024jlc, Mould:2025dts, Plunkett:2025mjr, Vitale:2025lms, Alvarez-Lopez:2025ltt, Wolfe:2025yxu}.
%
As we are solely interested in generating GW data with given component masses and distance,
we would like to estimate the detection probability
\begin{align}
    \p(\det\mid\mdone, \mdtwo, d_L)
    =
    \int \p(\det\mid\theta) \, \p(\widetildeparams{})
   \, \mathrm{d} \widetildeparams{}
     \,,
\end{align}
where $\theta=\{\mdone, \mdtwo, d_\mathrm{L},\widetilde{\theta}\}$, with $\widetilde{\theta}$ containing the source parameters in Table~\ref{tab: summary networks single-event prior} other than $\mdone$, $\mdtwo$, and $d_\mathrm{L}$, and $\p(\widetilde{\theta})$ the fixed single-event prior over $\widetilde{\theta}$.

We train a network to predict the detection probability; the architecture is summarized in Tab.~\ref{tab: architecture_summary}.
To optimize the network we use the binary cross-entropy loss, where the data are waveforms associated to $\mdone, \mdtwo, d_L$ that are labeled as either detected ($y_i = 1$) or undetected ($y_i=0$).
With $\estimatepdet$ the estimate of the network, the loss can be written as
\begin{align}
\nonumber
\text{Loss} 
= - \frac{1}{N}\sum_{i=1}^N \Big[
(1 - y_i)\log(1 - \hat{p}_i) + y_i \log\hat{p}_i
\Big]
\,,
\end{align}
with $N$ the batch size.
We follow standard \dingo training augmentation techniques (drawing the extrinsic parameters on the fly) and use as detection criteria a matched-filter SNR threshold of 12. We train for 200 epochs with a dataset containing $5\times 10^6$ waveforms. 
During training, the model thus iterates over $10^9$ GW signals. 

In Fig.~\ref{fig: pdet example}, we show the detection probability as a function of detector-frame primary mass and luminosity distance, fixing the secondary mass to $\mstwo=26~\msun$.

\subsection{Embedding emulator}

\begin{figure}
    \centering
    \includegraphics[width=\linewidth]{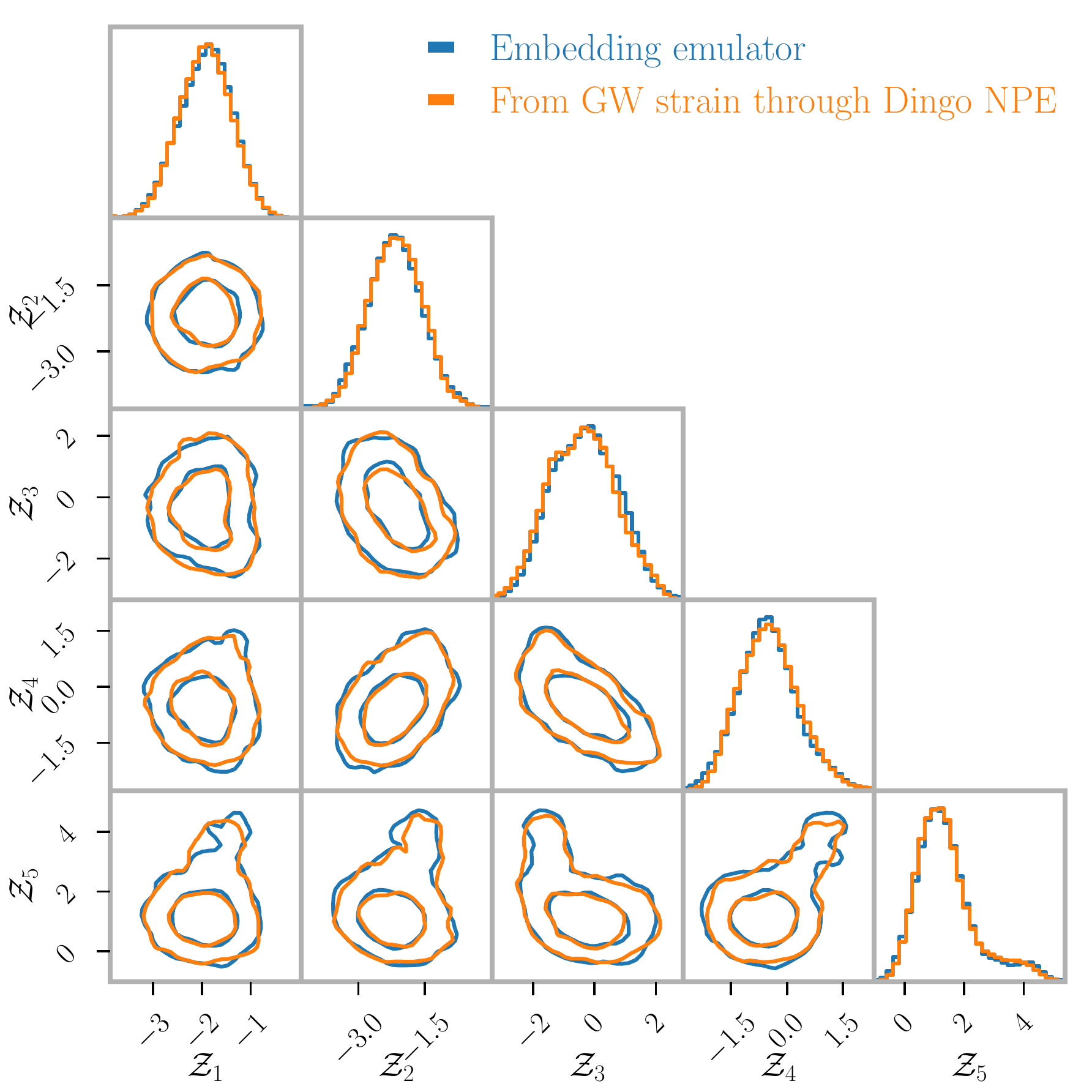}
    \caption{Distribution of embeddings in five out of the 32~embedding dimensions. For reference parameters of $\mathcal{M}_{\rm d} = 20.5$, mass ratio $q=0.8$ and luminosity distance $d_L = 1263$~Mpc we generate detected embeddings through the embedding emulator (blue) and by passing the GW strain data through the \dingo NPE network (orange). Plotted are the $0.5$ and $0.9$ levels. }
    \label{fig: validation embedding emulator}
\end{figure}

We train a normalizing flow to estimate $\p(\embedding\mid \mdone,\mdtwo,d_L,\det)$, as described in the main text; the architecture is summarized in Tab.~\ref{tab: architecture_summary}.
We sample source properties from the distribution in Tab.~\ref{tab: summary networks single-event prior}, generate synthetic waveforms with detector noise, and discard samples below our SNR detection threshold. The detected samples are passed through the single-event \dingo embedding network; conditioning the flow on $\mdone, \mdtwo, d_L$ marginalizes over the remaining source parameters and noise to give $\p(\embedding\mid \mdone, \mdtwo, d_L, \det)$.

After training the embedding emulator, we subsequently use it to directly sample event embeddings during training of the population transformer: parameters are drawn from the population model and accepted by the $p_\text{det}$ rejection step; for each accepted event, we sample $\embedding \sim \p(\embedding\mid \mdone,\mdtwo,d_L,\det)$ from the emulator. Examples are shown in Fig.~\ref{fig: validation embedding emulator}. This bypasses waveform generation and replaces it with fast sampling of $\embedding$.
For all tests of our trained model, we generate noisy waveforms and pass them through the original single-event \dingo embedding network, thus not relying on the embedding emulator.

\section{Validation}

\subsection{Validating individual network components}

\dingoP is based on three distinct models, each of which can be inaccurate:
(1) the estimator for detection probabilities given the detector-frame masses and luminosity distance;
(2) the emulator used only for training that produces detectable strain embeddings conditioned on the detector-frame component masses and luminosity distance;
and (3) the population transformer and normalizing flow that predicts the population posterior conditioned on the combined single-event embedding.
We test these after training to identify which (if any) is inaccurate.

The first test consists of computing a P--P plot using the detection estimator and embedding emulator to generate the data.
Since this data generation process is exactly the same as during training, this test validates whether the population transformer and the normalizing flow (3) reproduce the expected coverage  according to the hyperparameter posterior. 
Failing this test requires improving model (3).
This test provides no information about whether models (1) and (2) are accurate. 

If the above P--P test passes (using the embedding emulator and detection estimator) but then fails the P--P test described in the main text (using true waveforms that are then passed through the first part of the \dingo network to obtain the embeddings), this indicates that either model (1) or (2) are inaccurate. 
These models can be individually tested.

The samples from the embedding emulator can be compared to samples obtained by passing simulated signals added to detector noise through the single-event \dingo embedding network.
For the detection estimator, conventional Monte Carlo estimates (i.e., using injection studies, recovering simulated waveforms, and running detection pipelines) can be used to assess detection probabilities that can be readily compared. In practice, the embedding emulator is more likely to fail since the model is generative (hence much more complex), sampling a 32-dimensional distribution. 

\begin{figure*}[t]
    \centering
    \includegraphics[width=1.0\linewidth]{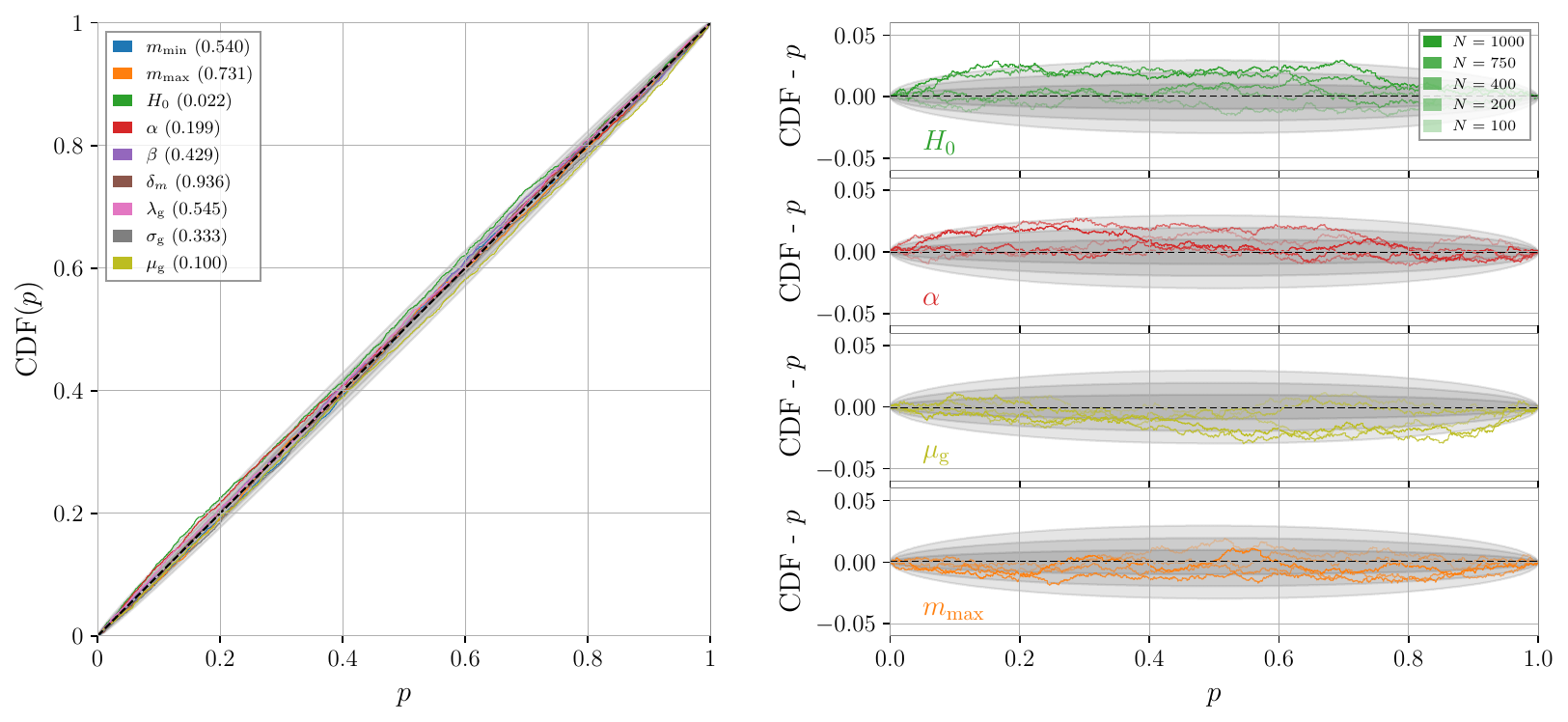}
    \caption{(\textit{Left}) P--P plot for posterior calibration for catalog size $N=1000$, with 2500 catalog realizations. 
    For each hyperparameter and simulated catalog, we compute the percentile rank of the true value within the marginal posterior. The cumulative distribution function (CDF) of these ranks is plotted; for well-calibrated posteriors, this follows the diagonal. Gray bands indicate 1-$\sigma$, 2-$\sigma$, and 3-$\sigma$ intervals expected under perfect calibration; Kolmogorov--Smirnov $p$-values are given in the legend. (\textit{Right}) For the four hyperparameters with the lowest $p$-values, the deviation $\mathrm{CDF}(p) - p$ from the diagonal is shown as a function of catalog size $N$. A curve consistently above (below) zero indicates the posterior is biased high (low) for that parameter; an ``S'' shape indicates over-dispersion.
    % For each $N$, we use $2500$ population realizations.
    }
    \label{fig: pp plot full}
\end{figure*}

For our trained networks, both variations of the P--P test pass, so the SBI approach is internally consistent: model (3) has been correctly trained, and the emulator and detection estimator are working to sufficient accuracy. The emulator-based variant is consistent with the full-pipeline variant and is not shown for brevity.
The left panel of Fig.~\ref{fig: pp plot full} shows the full-pipeline P--P test for 2500 realizations of a GW catalog of size 1000. Compared to Fig.~\ref{fig: pp plot} (with $N \sim \mathcal{U}(25, 1000)$), this tests the model at the upper end of its training range, where sharper posteriors make the test more sensitive to residual biases; the lower $p$-values reflect both this increased sensitivity and degradation at the training-set boundary.
The right panel of Fig.~\ref{fig: pp plot full} shows the P--P test as a function of catalog size. While $H_0$, $\alpha$, and $\mmax$ show weak $N$ dependence, the $\mug$ posterior approximation worsens with increasing $N$. 

A final concern is that the 32-dimensional embeddings might be insufficient to capture all information from each BBH signal needed for population inference. In this case, the \dingoP posterior would be broader than the conventional result (without being biased) even when both P--P variants pass, since information has been lost. This is addressed by direct comparison to the conventional likelihood-based result.

\subsection{Comparison to standard HBA}

%The following section provides comparisons between our SBI approach and the conventional approach that is likelihood-based and uses single-event posterior samples. 
For each of the two populations defined by the hyperparameter sets of Tab.~\ref{tab:HBAcomparison}, we draw 500 events and corresponding strain data.
With \textsc{Dingo-IS}, we produce single-event posterior samples with which we perform conventional HBA with \icarogw{}.
As 56 (20) of these events have low effective sample size for Population 1 (2), we run these events with \textsc{Bilby} to avoid biased single-event posterior samples. 
We compute the population-level selection function using $3.8\times 10^6$ sources that pass our SNR threshold. 

% \begin{table}[h]
% \caption{Hyperparameters assumed for the verification and application datasets. The first two columns correspond to validation populations, while the last column corresponds to the $H_0$ uncertainty application discussed in the text.
% \\\mm{I'm confused about what this table is showing. Are the columns the true hyperparameters? I thought for the relative $H_0$ uncertainty tests that various populations (different hyperparameters) were used? Or is the final column just for the 2D figure, not the one in the main text?} 
% }
% \label{tab:hyperparameters training sets verification}
% \begin{ruledtabular}
% \begin{tabular}{lccc}
% Hyperparameter & \multicolumn{2}{c}{Validation} & Application \\
% %\cline{2-3} 
% & \#1 & \#2 & $H_0$ uncert. \\
% \colrule
% $m_{\min} \ [M_\odot]$ & 18.14 & 21.65 & 21.0 \\
% $m_{\max} \ [M_\odot]$ & 40.76 & 38.38 & 41.05 \\
% $H_0 \ [\mathrm{km\,s^{-1}\,Mpc^{-1}}]$ & 64.71 & 72.29 & 67.0 \\
% $\alpha$ & 1.11 & -0.39 & 2.0 \\
% $\beta$ & -1.06 & 2.29 & 1.0 \\
% $\delta_{m} \ [M_\odot]$ & 4.75 & 3.78 & 4.1 \\
% $\lambda_{\mathrm{g}}$ & 0.02 & 0.67 & vary \\
% $\sigma_{g} \ [M_\odot]$ & 5.99 & 7.62 & vary \\
% $\mu_{g} [\ M_\odot]$ & 24.98 & 21.10 & 28.2 \\
% \end{tabular}
% \end{ruledtabular}
% \end{table}

\begin{figure*}[!p]
    \centering
    \includegraphics[width=\linewidth]{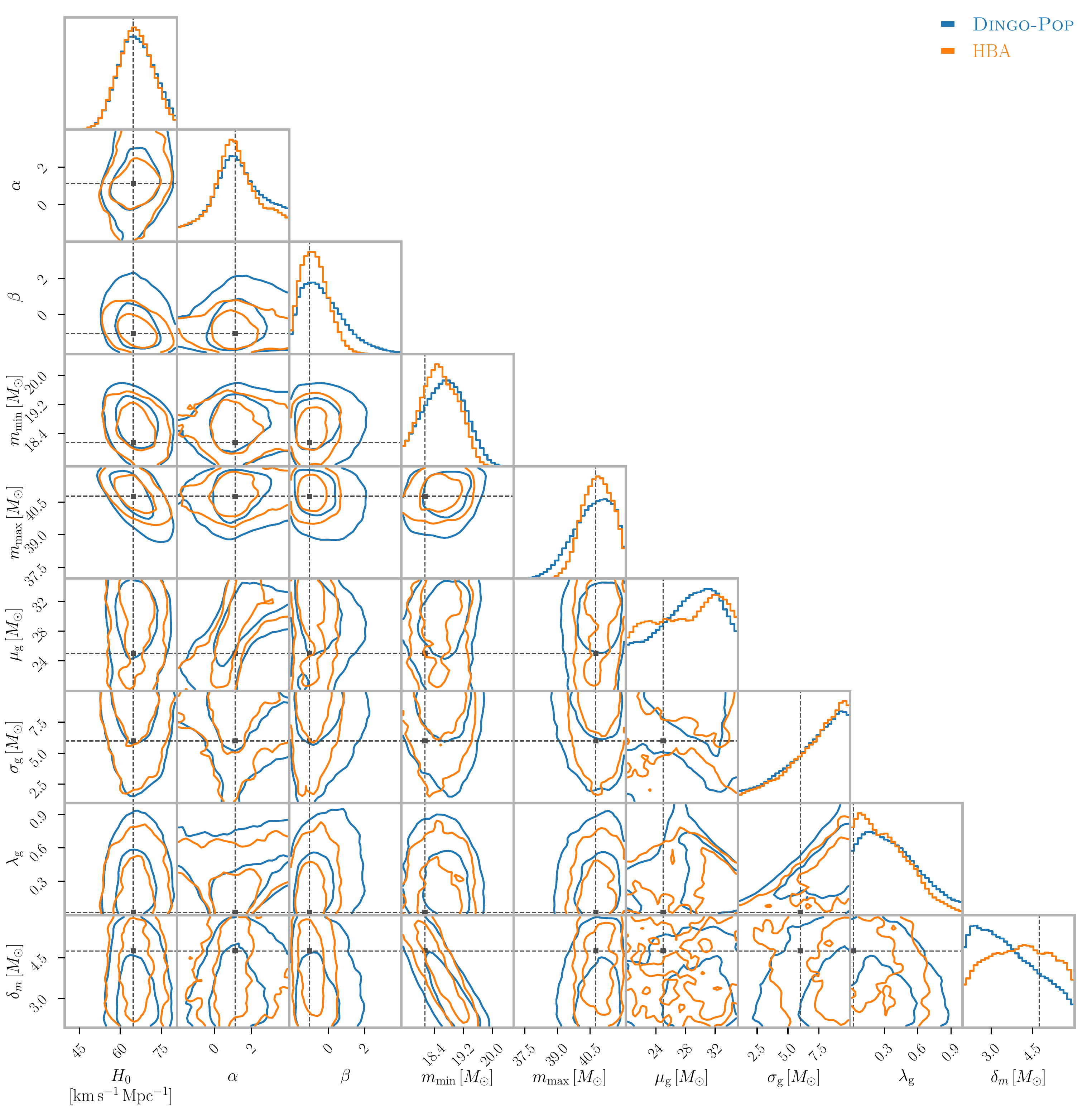}
\caption{Comparison between the conventional likelihood-based (orange) and likelihood-free SBI (blue) methods for Population~1 (cf~Tab.~\ref{tab:HBAcomparison}) with 500 events analyzed.
    The two methods agree well. 
    The true hyperparameters are marked with gray solid lines. 
    Also indicated are the 50\% and 90\% credible regions.
    }
    \label{fig: comparison likelihood-based result}
\end{figure*}

\begin{figure*}[!p]
    \centering
    \includegraphics[width=\linewidth]{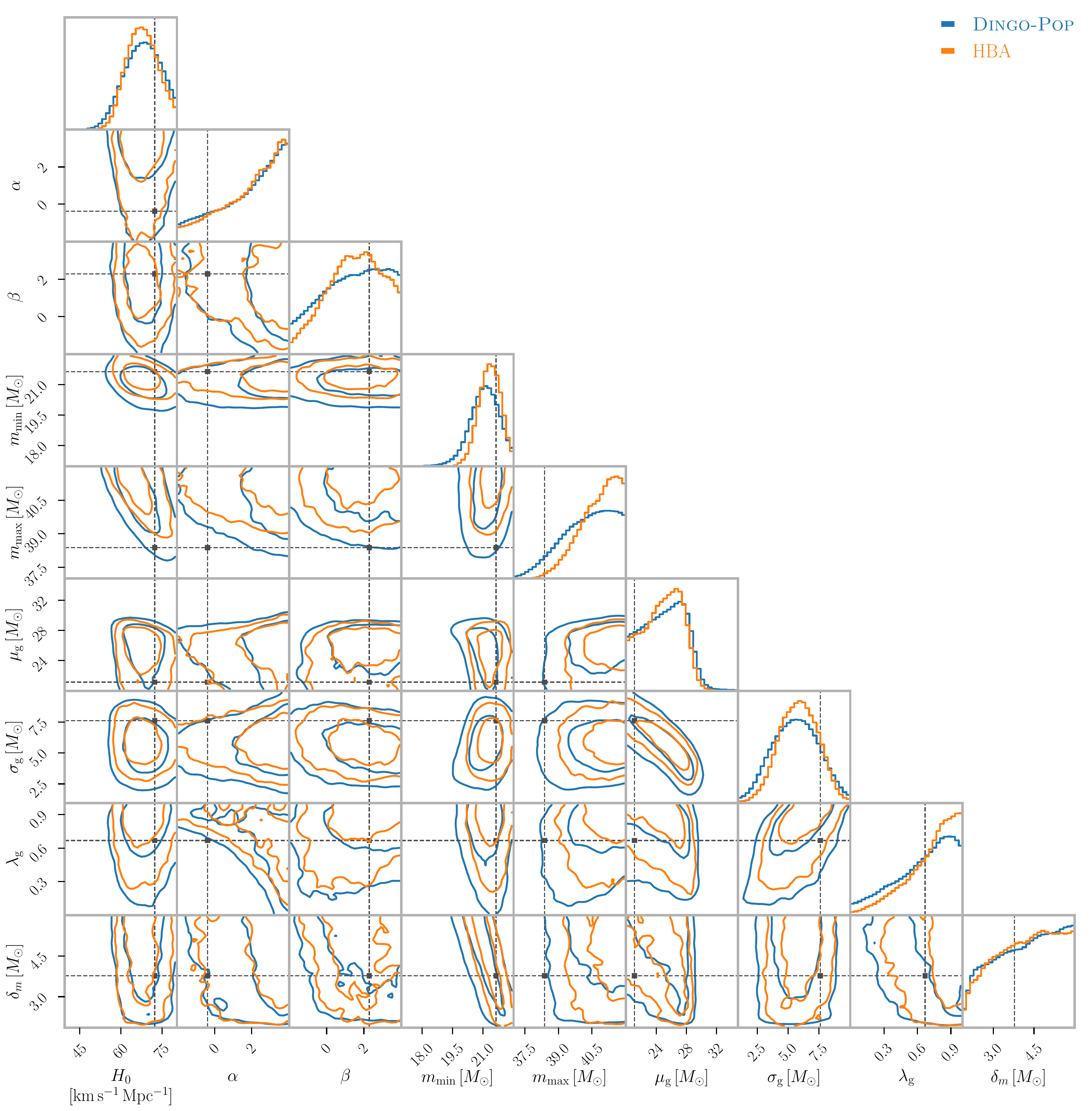}
    \caption{As in Fig.~\ref{fig: comparison likelihood-based result} but for Population~2.}
    \label{fig: comparison likelihood-based result appendix}
\end{figure*}

Figures~\ref{fig: comparison likelihood-based result} and \ref{fig: comparison likelihood-based result appendix} compare the conventional and \textsc{Dingo-Pop} results for the hyperparameter posterior.
% for a catalog of size 500 for populations~1 and 2, respectively. 
They agree well, with the largest discrepancies in the hyperparameters $\beta, \delta_m$, $\mmax$, and $\mug$.
There are several possible explanations for this discrepancy:
(1) the SBI posterior may be biased, but the P--P test of Fig.~\ref{fig: pp plot} passes, which makes this scenario less likely;
(2) the event embeddings may not be expressive enough, and while this would not necessarily entail a bias, it should lead to a broadening of the \dingoP posterior;
(3) the Monte Carlo uncertainties of the conventional HBA approach may not be under control. We perform a sensitivity analysis to test this last possibility.

We note that the SBI approach is closer to the true hyperparameter values for the minimum and maximum masses $m_\mathrm{min}$ and $m_\mathrm{max}$ in Population 2, suggesting that the SBI posterior is more robust than the likelihood-based result.
Since these parameters represent the boundary of the population, they are more affected than the remaining parameters by the Monte Carlo uncertainty from the single-event posteriors.

% \begin{table}[htbp]
% \centering
% \renewcommand{\arraystretch}{1.7}
% \begin{tabular}{|c|c|c|c|}
% \hline
% $\hyperparams$ & \multicolumn{2}{c|}{\textbf{Dataset}} & \textbf{Unit} \\
% \hline
% & 0 & 1 & \\
% \hline
% $H_0$ & 64.71 & 91.29 & $\mathrm{km\,s^{-1}\,Mpc^{-1}}$ \\
% $\mmin$ & 18.14 & 20.49 & $\msun$ \\
% $\mmax$ & 40.76 & 44.42 & $\msun$ \\
% $\alpha$ & 1.11 & 0.51 & - \\
% $\beta$ & -1.06 & -0.27 & - \\
% $\delta_{\rm m}$ & 4.75 & 3.35 & $\msun$ \\
% $\lambda_{\mathrm{peak}}$ & 0.02 & 0.56 & - \\
% $\mu_{\rm g}$ & 24.98 & 31.08 & $\msun$ \\
% $\sigma_{\rm g}$ & 5.99 & 6.55 & $\msun$ \\
% \hline
% \end{tabular}
% \caption{Hyperparameters assumed for the verification dataset. For each population we analyze \mightchange{600} GW events and produce single-event posterior samples needed for the likelihood-based analysis. }
% \label{tab:hyperparameters training sets verification}
% \end{table}

\subsection{Sensitivity of HBA to Monte Carlo approximations}

% Various approximations underpin the likelihood-based approach. The most important are the approximation of the integrals appearing in Eq.~\eqref{eq:pop-likelihood} through single-event posterior samples and the injection set. 
To ensure that the analysis results are independent of the number of single-event posterior samples, we repeat the analysis twice with different PE sample sizes. A comparison of results is given in Fig.~\ref{fig:impact number posterior samples, 500 events}. There is no clear change with the number of samples per event.

\begin{figure*}[!p]
    \centering
    \includegraphics[width=1\linewidth]{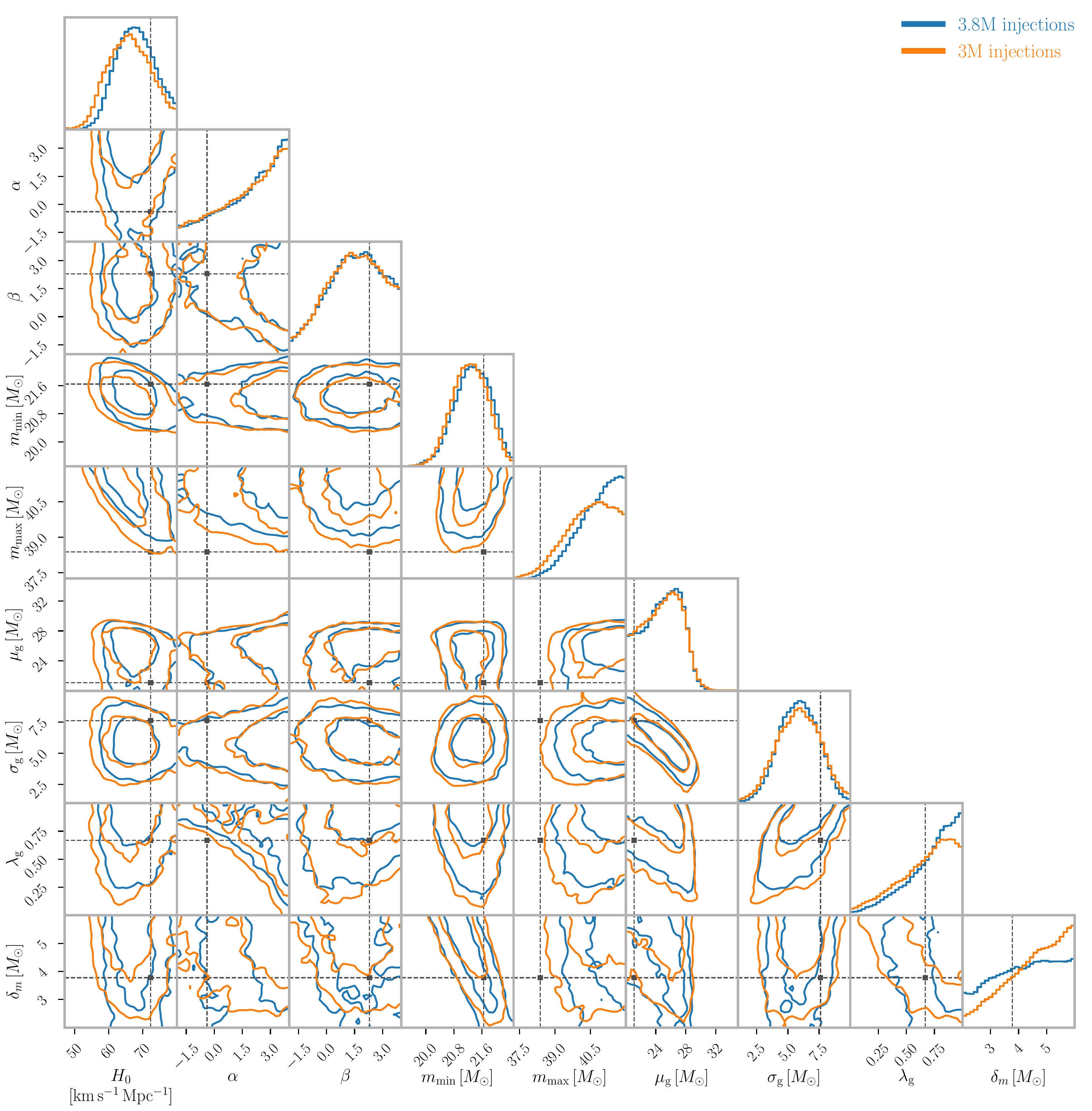}
    \caption{Conventional analysis for Population 2, comparing the hyperparameter posterior for different numbers of injections to estimate selection effects, as indicated in the legend. The catalog contains 500 events with 4096 samples each. Panels with marginal two-dimensional posteriors show 50\% and 90\% credible regions. True parameter values are marked with black lines.}
    \label{fig:impact number injection sets}
\end{figure*}

Similarly, in Fig.~\ref{fig:impact number injection sets}, we test for dependence on the number of signals contained in the injection set. While most hyperparameters show a weak dependence, there are several exceptions: $\mmax$, $\delta_m$, and $\lambda_{\rm g}$, indicating that these parameters are particularly vulnerable to the Monte Carlo uncertainty.
The Monte Carlo uncertainty related to the injection set is non-negligible
% (e.g.~in $\delta_m$,
and leads to scatter in the hyperparameter posterior.
These deviations are comparable to the differences with \dingoP, indicating that injection-set MC uncertainty is the dominant contributor to the \dingoP–HBA discrepancy.
However, the SBI posterior is still broader in $\beta$, a parameter which appears to be robust to the number of injection samples.

\subsection{Out-of-distribution datasets}

SBI results can be unpredictable in case of model misspecification.
In this section we test how \textsc{Dingo-Pop} performs on datasets from a different population model than the one assumed during training. 
As an out-of-distribution dataset, we generate a population from the \plp model but create a gap in the primary mass.
We first generate 200 events from the \plp with 
%the hyperparameter values reported in Tab.~\ref{tab:ood} 
$m_{\rm min} = 18~\msun$, $m_{\rm max} = 40~\msun$, $H_0 = 70~\text{km}\,\text{s}^{-1}\,\text{Mpc}^{-1}$, $\alpha=3$, $\beta = 1$, $\delta_{m} = 4~\msun$, $\lambda_{g}=0.5$, $\sigma_{g} = 2~\msun$, and $\mu_{g} = 30~\msun$. We then remove the 100 events with primary mass closest to $27~\msun$.
% The resulting population has 100 events.
%We choose a combination of hyperparameters and mass gap location to produce a dataset unlikely to arise as a fluctuation of the \plp model during training.
Under the assumption of the incorrect \plp mass model, we then analyze the out-of-distribution dataset using both the conventional method and SBI.
For the conventional analysis, we use $4\times 10^{5}$ injections and 5000 PE samples generated using a \textsc{Dingo} GNPE network~\cite{Dax:2021tsq} with importance sampling.
No thresholds are applied to the Monte Carlo variance of the likelihood. 

\begin{figure*}[!p]
    \centering
    \includegraphics[width=1\linewidth]{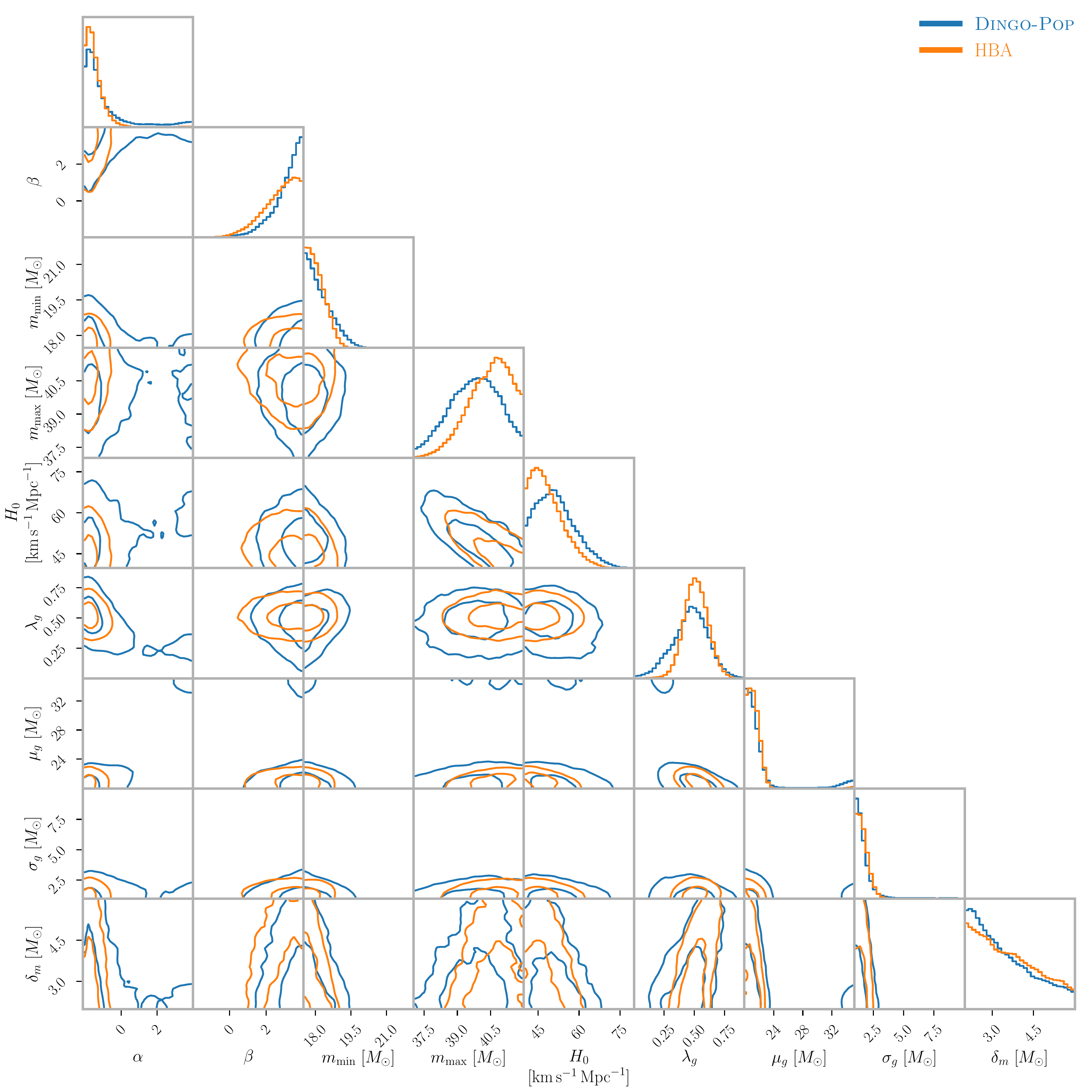}%
    \llap{\raisebox{10.5cm}{\includegraphics[width=0.45\linewidth]{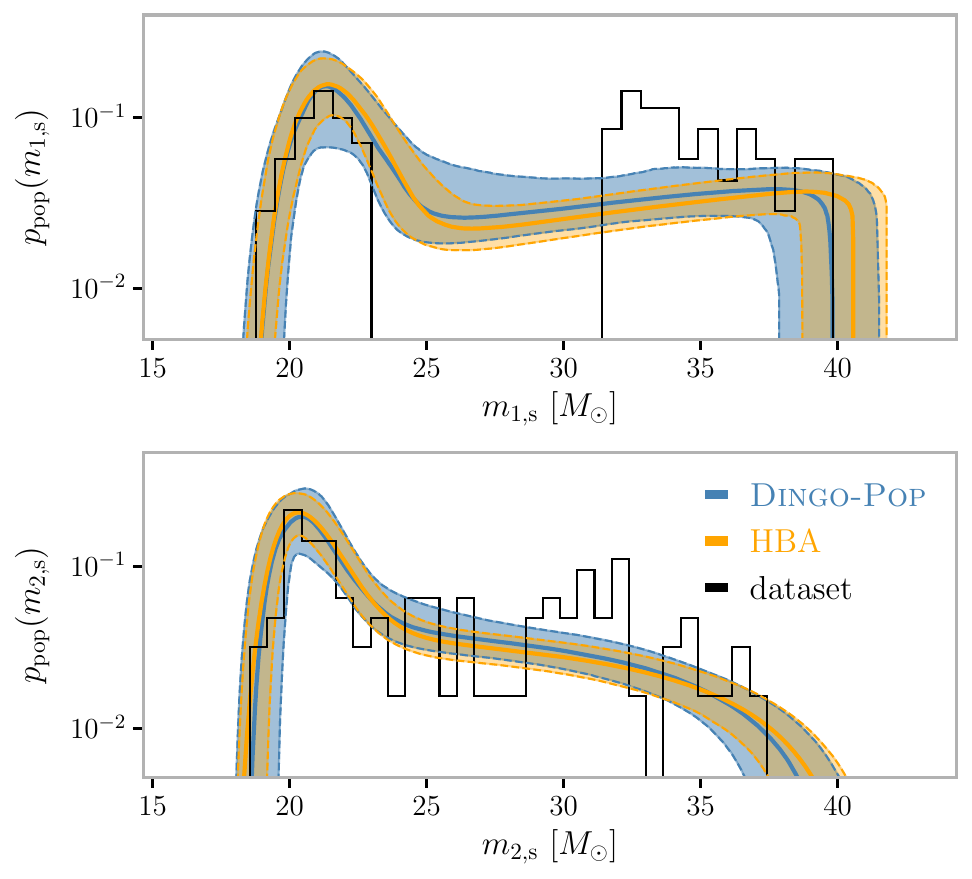}}}
    \caption{Comparison for the 100-event out-of-distribution catalog (cf. Tab.~\ref{tab:ood}) between the conventional likelihood-based (green) and likelihood-free SBI (blue) methods. 
    In the corner plot, contours contain 50\% and 90\% credible regions. The top-right panels show the posterior medians and 90\% credible regions of the inferred source population over primary and secondary masses. The black histogram shows the distribution of events in the simulated catalog (not the underlying astrophysical population).
    %\kcom{Can we make the legend larger? Similar to the above corner plots? }
    }
    \label{fig:ood}
\end{figure*}

In Fig.~\ref{fig:ood}, the corner plot and mass spectra show that the results from \dingoP are broadly consistent with the conventional analysis.
% \cf{\sout{, with marginalized posteriors generally broader for \dingoP}}.
\dingoP exhibits a weak bimodality on the right tail of $\alpha$ and $\mu_{g}$ that is absent from the conventional analysis. The presence of these extended tails results in broader credible intervals in the \dingoP mass spectra, particularly for the primary mass.
Table~\ref{tab:ood} reports medians and $90\%$ credible intervals for both analyses.

\begin{table}[h]
\caption{
Comparisons of \textsc{Dingo-Pop} and conventional HBA approaches applied to an out-of-distribution population. The hyperparameters are used to generate a 200-event catalog from the \plp model. We then make it out-of-distribution by creating a gap in the primary mass, removing the 100 events with primary mass closest to $27~\msun$.
}
\centering
\setlength{\tabcolsep}{13pt}
\renewcommand{\arraystretch}{1.2}
\begin{ruledtabular}
\begin{tabular}{l c c}
  % & \multicolumn{2}{c}{Out-of-distribution population} \\
Hyperparameters $\Lambda$ & \dingoP & HBA \\
\hline
$m_{\rm min}$ $[\msun]$ & 18.0 $^{+1.1}_{- 0.5}$ & 18.0 $^{+0.7}_{-0.4}$ \\
$m_{\rm max}$ $[\msun]$ & 39.8 $^{+1.7}_{-2.0}$ & 40.6 $^{+1.2}_{-1.9}$\\
$H_0$ $[\text{km}\,\text{s}^{-1}\,\text{Mpc}^{-1}]$ & 50.8 $^{+14.1}_{-9.3}$ & 47.3 $^{+12.3}_{-6.4}$\\
$\alpha$ & $-1.5$ $^{+4.4}_{-0.5}$ & $-1.7$ $^{+1.0}_{-0.3}$\\
$\beta$ & 3.3 $^{+0.6 }_{-1.8}$ & 2.9 $^{+1.0}_{-2.0}$\\
$\delta_m$ $[\msun]$ & 3.2 $^{+2.3}_{-1.1}$ & 3.4 $^{+2.2}_{-1.3}$\\
$\lambda_{\rm g}$ & 0.5 $^{+0.2}_{-0.3}$ & 0.5 $^{+0.2}_{-0.2}$\\
$\sigma_{\rm g}$ $[\msun]$ & 1.4 $^{+ 2.1}_{-0.4}$ & 1.5 $^{+0.9}_{-0.5}$\\
$\mu_{\rm g}$ $[\msun]$ & 21.0 $^{+12.5}_{-0.4}$ & 21.1 $^{+1.3}_{-1.0}$\\
\end{tabular}
\end{ruledtabular}
\label{tab:ood}
\end{table}

\begin{figure*}[!p]
    \centering
    \includegraphics[width=1\linewidth]{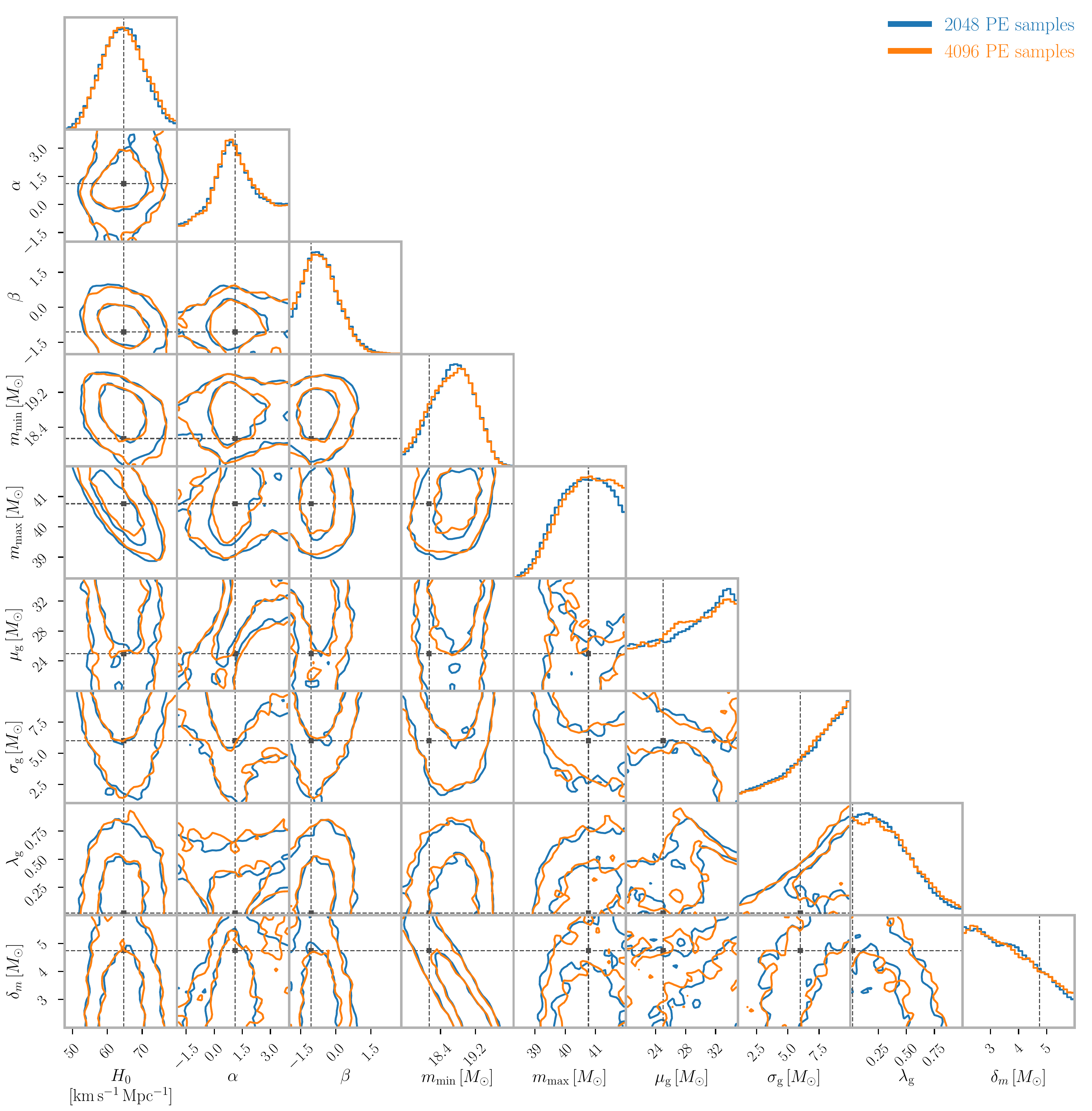}
    \caption{Conventional analysis for Population 1, comparing the hyperparameter posterior for different numbers of single-event samples, as indicated in the legend. The catalog contains 500 events and $8\times10^5$ injections are used to compute the selection effect. Panels with marginal two-dimensional posteriors show 50\% and 90\% credible regions. True parameter values are marked with black lines.}
    \label{fig:impact number posterior samples, 500 events}
\end{figure*}

\section{Additional results}

\begin{figure*}
    \centering
    \includegraphics[width=1\linewidth]{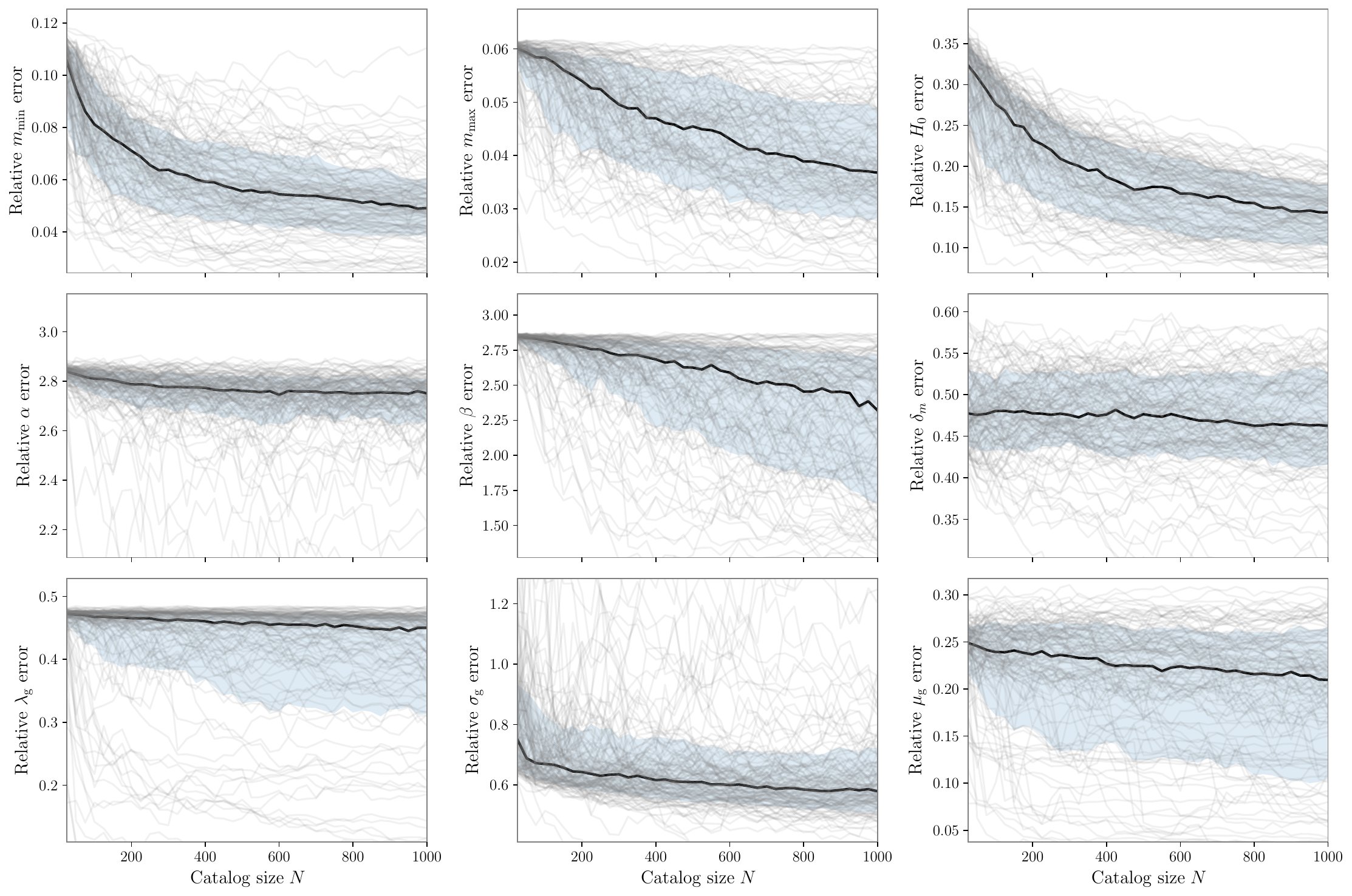}
    \caption{
    Repetition of Fig.~\ref{fig:examples_applications} but for all hyperparameters. To reduce fluctuations, for $\alpha$, $\beta$, and $\lambda_{\rm g}$ we use only the 2-$\sigma$ interval (not dividing by the median as for the rest of the parameters).}
    \label{fig: evolution relative uncertainty all params}
\end{figure*}

We plot the relative uncertainties of all inferred hyperparameters as a function of the catalog size in Fig.~\ref{fig: evolution relative uncertainty all params}.
The parameters $\mmin$, $\mmax$, $H_0$, $\beta$, $\sigmag$, and $\mug$ show the expected trend of decreased uncertainty with increasing catalog size, but the parameters $\alpha$, $\delta_m$, and $\lambda_{\rm g}$ do not show this trend.
For these latter variables, the uncertainties decrease for some population realizations (blue, semi-transparent lines), while the median is almost constant, indicating that for most realizations the result is prior driven.

% Table~\ref{tab:HBAcomparison} summarizes the hyperparameters associated to the scaling applications shown in Fig.~\ref{fig: examples applications}.

% \begin{figure}[t]
%     \centering
%     % \vspace{1.2cm}
%     \includegraphics[width=1.0\linewidth]{Delta_H0_with_lambda_peak_sigma_g.pdf}
%     \caption{The $H_0$ uncertainty (1-sigma interval) as a function of the fraction of events in the Gaussian component $\lambda_\mathrm{g}$ and the width of the Gaussian component $\sigma_\mathrm{g}$.
%     The other hyperparameters are kept fixed as given in Tab.~\ref{tab:hyperparameters training sets verification}.
%     Each point is averaged over five population realizations, each with 400 detected events.
%     }
%     \label{fig:examples_applications-2}
% \end{figure}

% Create the reference section using BibTeX:

\end{document}